\documentclass[
  aps,
  prb,
  twoside,
  twocolumn,
  floatfix
]{revtex4}
\usepackage{amsmath}
\usepackage{amssymb}
\usepackage{graphicx}
\usepackage{tabularx}
\usepackage{dcolumn}
\usepackage{longtable}
\usepackage{footnote}
\usepackage{appendix}

\begin{document}

\newcommand{\bec}{\begin{center}}
\newcommand{\ec}{\end{center}}
\newcommand{\be}{\begin{equation}}
\newcommand{\ee}{\end{equation}}
\newcommand{\beqn}{\begin{eqnarray}}
\newcommand{\eeqn}{\end{eqnarray}}
\newcommand{\bet}{\begin{table}}
\newcommand{\ent}{\end{table}}
\newcommand{\bib}{\bibitem}

% commands for structuring
\newcommand{\sect}[1]{Sect.~\ref{#1}}
\newcommand{\fig}[1]{Fig.~\ref{#1}}
\newcommand{\Eq}[1]{Eq.~(\ref{#1})}
\newcommand{\eq}[1]{(\ref{#1})}
\newcommand{\tab}[1]{Table~\ref{#1}}

% commands for mathmode
\renewcommand{\vec}[1]{\ensuremath\boldsymbol{#1}}
\renewcommand{\epsilon}[0]{\varepsilon}

% comments at the margin
\newcommand{\cmt}[1]{\emph{\color{red}#1}%
  \marginpar{{\color{red}\bfseries $!!$}}}

%\topmargin 0.5cm
%\setlength{\textwidth}{18cm} 
%\setlength{\textheight}{25.5cm} 

%\baselineskip 4mm 
%\wideabs{

\title{
%\LARGE{
%Graphene moir\'{e} superlattices on Cu(111) and parameter fitting for
%classical molecular dynamics simulations 
Time-lapsed graphene moir\'{e} superlattice on Cu(111) 
%Dynamical graphene moir\'{e} superlattice on Cu(111) 
}

%\LARGE{

\author{P. S\"ule}
\affiliation{Research Centre for Natural Sciences,
Institute for Technical Physics and Materials Science
Konkoly Thege u. 29-33, Budapest, Hungary}

\author{M. Szendr\H{o}}
\affiliation{Research Centre for Natural Sciences,
Institute for Technical Physics and Materials Science
Konkoly Thege u. 29-33, Budapest, Hungary}

\date{\today}

\begin{abstract}

 The detailed study of the graphene (gr) moir\'{e} superlattices emerging due to the mismatch between the substrate's and gr-overlayer crystal lattices is inevitable because of its high technological relevance. However, little is known about the dynamics of moir\'{e} superstructures on gr.
Here, we report the first classical molecular dynamics simulation (CMD) of the moir\'{e} superlattice of graphene on Cu(111) using a new parameterized Abell-Tersoff-potential for the graphene/Cu(111) interface fitted in this paper to nonlocal van der Waals density functional theory (DFT) calculations. The interfacial force field with time-lapsed CMD provides superlattices in good quantitative agreement with the available experimental results. The long range coincidence supercells
of $2 \times 2$ and $3 \times 3$ with nonequivalent moir\'{e} hills have also been identified and analyzed.
The moir\'{e} superlattice exhibits a pattern which is dynamical
rather than statically pinned to the support and can be observed
mostly via time lapsing. The 
instantaneous snapshots of the periodic moir\'{e} pattern
already at low temperature are weakly disordered lacking the apparent sharpness of the time averaged pattern and scanning tunneling
microscopy images.
This suggests the existence of competing orders between a static (1st order) and
a dynamical (2nd order) moir\'{e} superstructures.
The revealed random height fluctuations may limit
the important electronic properties of supported graphene such as
the mobility of charge carriers. 
\\
%{\em keywords}: graphene, moire patterns, superlattices, atomistic and nanoscale simulations, molecular dynamics simulations
%\pacs{61.48.Gh,68.35.B-,83.10.Rs,31.15.A-}
%}

\end{abstract}

%}

\maketitle

%\scriptsize{
%graphene structure, 61.48.Gh
%Surface reconstruction, 68.35.B-
%computer simulation of, 83.10.Rs
%%%%%%%%%%%%%%%%%%%%%%%%%%%%%%%%%%%

%%%%%%%%%%%%%%%%%%%%%%%%%%%%%%%%%%%%%%%%%%%%%%%%%%%%%%%%%%%%%%%%%%%

\footnotetext[1]{Corresponding author, E-mail: sule@mfa.kfki.hu (P\'eter S\"ule)}

\section{Introduction}

 Graphene is a subject of intensive and booming research efforts either in its suspended and 
supported forms \cite{Novoselov,Neto,Meyer,Tapaszto}.
 Understanding the properties of the interface
between graphene and a metal support has gained considerable
attention due to the fact that graphene/metal contacts could be required by
future nanoscale electronics \cite{Crommie}.

 Graphene (gr) when placed on a substrate becomes periodically patterned induced by the lattice misfit
 between the substrate lattice and the overlayer \cite{Batzill,rev}. The emerging long-range periodic moir\'{e} nano-superlattice
 will be influenced mostly by the local binding environment. The small group of Carbon atoms will rise with respect to the substrate
 when locked-in in a hollow position (the first neighbor substrate atom is in the middle of a Carbon hexagon, hollow humps or protrusions) and other group of atoms will get closer to the
 substrate which bind to ontop positions (support atoms are nearly covered by Carbon atoms, ontop bumps or bulged-in regions). The alternating arrays of such regions build up the moir\'{e} superlattice
 with different periodicity and height variation on different supports \cite{DFT:Ru-Stradi,Stradi13,DFT:Ru-Hutter}. Depending on the strength of the gr-support
 interaction, the height measured from the deepest point of the bump region to the peak of the hump (bump-to-hump corrugation) can be substantial reaching 0.1 nm \cite{Batzill,rev}. Four moir\'{e} hills (called moirons \cite{grru13}) form a minimal rhombic supercell centered at the corner points of the rhombus. 
When these moirons are equivalent, the system can be described
by a relatively simple supercell which includes one moiron \cite{DFT:Ru-Hutter,grru13}.

 However, it turned out,
 that a larger coincidence superstructure is the "real" supercell of the gr/Ru(0001) system in which the
 four humps are translationally and structurally inequivalent \cite{grru13,SXRD:25X25,rotRu,Parga}.
 The situation is even more complicated because 
 the periodicity of the moir\'{e} superlattice further depends on the rotation misalignment of the gr-sheet with respect to the
 support sheet \cite{Gao,JACS-grcu,grcu:LEEM,rotPt,rotRu,rotIr2}.
%Multi-oriented rotational moir\'{e} superstructures of graphene on Pt(111) and Ir(111)
%have also been found \cite{rotPt,rotIr2,rotPt2,rotIr3}.
%Supported graphene exhibits a large conincidence superlattice which is the large unit cell
%of moir\'{e} superstructures. 

 The detailed study of various graphene (gr) superlattices,
such as the moir\'{e} superstructures \cite{Batzill,rev} and the corresponding nanoscale topography 
requires sophisticated experimental and theoretical methods \cite{DFT:Ru-Stradi,Stradi13,DFT:Ru-Hutter,grru13}.
  The theoretical modelling of weak adherence of gr to the support is essential to analyse and verify the experimental
results. However,  {\em ab initio} density functional theory (DFT) calculations and geometry optimization
can be carried out for systems with few thousand atoms \cite{grru13}. Above 1000 atoms in general, however,
electronic structure calculations become difficult even on top-level supercomputers. Therefore, it is important
to find more efficient approaches which can handle routinly large scale systems. Classical molecular dynamics (CMD)
simulations offer a powerfull tool for the structural and energetic analysis of gr-systems \cite{Sule13}.
The problematic part of such simulations is the interfacial interaction between the gr sheet and support.  
While few reliable empirical potentials are available for graphene-only simulations
\cite{airebo,lcbop},
however,
little attention has been paid to the adequate description of gr/support
interfaces.
The weak adherence, the site selected binding of gr on various supports
and the periodic moir\'{e} topography requires sophisticated modelling
which goes beyond the level of simple pairwise force fields \cite{Sule13}. 

 It has been shown recently that the development of a new angular dependent bond order interfacial
force field provides the adequate description of the prototypical gr/Ru(0001)
system \cite{Sule13}.
Rotation misorientated moir\'{e} superlattices have also been
studied recently and a new phase has been explored by STM and CMD simulations
\cite{Sule14}.
 First principles calculations (such as DFT) have widely been
used in the last few years to understand corrugation of nanoscale gr sheets
on various substrates \cite{DFT:Ru-Hutter,DFT:Ru-Stradi,DFT:Ru_Wang,DFT:Ru_corrug},
modelling larger systems, above $1000$ Carbon atoms
remains, however challenging, especially if geometry optimization
is included and/or large supercells are considered \cite{grru13}. 
 The minimal supercell of the gr/Cu(111) system
includes a few thousands of atoms which definitely
exceeds the size limit of accurate DFT geometry optimizations
and/or {\em ab initio} DFT molecular dynamics.

 Here we show that using a new
DFT adaptively parameterized interfacial Abell-Tersoff (AT) potential \cite{Abell,tersoff}
one can quantitatively reproduce even the fine structure of the experimentally observed surface reconstructions of gr on Cu(111) (moir\'{e} superstructures).
  Moreover, the dynamic nature of the periodic moir\'{e} pattern 
has also been revealed which has not yet fully been realized until now.

\section{Methodology}

%%%%%%%%%%%%%%%%%%%m
\subsection{Simulation rules}
%%%%%%%%%%%%%%%%%%%

 Classical molecular dynamics has been used as implemented
in the LAMMPS code (Large-scale Atomic/Molecular Massively Parallel Simulator) \cite{lammps}.
The graphene layer has been placed nearly commensurately on the substrate
since the lattice mismatch is small in gr/Cu(111).
However, even this tiny misfit is sufficient
to form an incommensurate overlayer by occupying
partly registered positions (alternating hexagonal hollow and ontop sites) which
leads to a moir$\acute{e}$ superstructure.

 Periodic triclinic (rhombic) simulations cells have been constructed between $85 \times 85$
and $255 \times 255$ gr-unit cells.
%We find these structures to be suitable for simulations leading to
%nearly commensurate superstructures.
The systems are carefully matched at the cell borders in order to 
give rise to perfect periodic systems.
Arbitrary system sizes lead to non-perfect matching at the cell borders.
Moreover,
nonperiodic cells lead to unstable moir\'{e} patterns due to the
undercoordinated atoms at the system border which cannot be handled
by the present force field with CMD.
Nonperiodic structures can be, however, optimized by simple
minimizers which also provide moir\'{e} patterns.
%This pattern sometimes is not sufficiently developed and becomes
%unstable during nonperiodic CMD due to the
%improper treatment of the undercoordinated border atoms.
Further refinement of the periodic pattern can be obtained by
time-lapsed CMD.
The moir\'{e} pattern is extremely sensitive to weak effects during CMD such as 
e.g. the improperly treated border atoms and/or the arising tensile stress or strain at the simulation cell border.

 Isobaric-isothermal (NPT ensemble) simulations (with Nose-Hoover thermostat and a prestostat) were carried out at 0-300 K. Vacuum regions were inserted
above and below the slab of the gr-substrate system to ensure
the periodic conditions not only in lateral directions (x,y) but also in the direction perpendicular to the gr sheet (z).
The variable time step algorithm has been exploited.
The codes OVITO \cite{ovito} and Gnuplot
have been utilized for displaying atomic and nanoscale structures \cite{Sule13,Sule_2011}.

 The molecular dynamics simulations allow the optimal lateral
positioning of the gr layer in order to reach nearly epitaxial displacement
and the minimization of lattice misfit.
The relaxation of the systems has been reached in 2 steps:
first conjugate gradient geometry optimization (cg-min) in combination of simulation box relaxation (boxrel) of the rhomboid simulation cell has been carried out.
Finally variable time step CMD simulations have been utilized in
few tens of a thousand simulation steps to allow the further reorganization
of the system under thermal and pressure controll (NPT, Nose-Hoover
thermostat, prestostat).
Therefore we use in general the combined cg-min/CMD simulations.

%{\em Frequency of moir\'{e} patterning and time-lapsing:}
  Time-lapsed CMD (TL-CMD) has been used at 0-300 K in order to average the morphology over longer timescale and also to account for the effect of temperature. We found that 10000 time steps are generally sufficient for a stable moir\'{e}
pattern. The pattern remains stable for time averages of much longer simulations.
Time-lapsing is important because the moir\'{e} pattern becomes much less sharp
for short simulations (for less than 5000-10000 steps) or for single-time-points.
This is because sharp moir\'{e} patterning seems to appear beyond a certain timeperiod 
being a dynamic phenomenon rather than a static one in gr/Cu(111).
We estimate the frequency of the occurrence of a sharp moir\'{e} superlattice $\nu \approx 1/\tau \approx 10^{12}$ Hz with the periodic time of $\tau \approx
1$ ps. 
%The dynamical moir\'{e} pattern is
%easily destroyed by atomic vibrations in a short time scale.

 The AIREBO (Adaptive Intermolecular Reactive Empirical Bond Order) 
potential
has been used for the graphene sheet \cite{airebo}.
The more recently developed long-range bond-order potential for carbon (LCBOP) has also been
employed for comparative purpose \cite{lcbop}, although not much difference
have been found in the essential properties, therefore we do not show
explicit results on that potential.
For the Cu substrate, a recent embedded atomic method (EAM) \cite{EAM:Cu} potential is employed.

 For the C-Cu interaction we developed a new Abell-Tersoff-like 
angular-dependent potential \cite{Abell,tersoff} (see Supplementary Material for further details).
In the AT potential file (lammps format) the C-C and Cu-Cu interactions are ignored (nulled out).
The CCuC and CuCCu out-of-plane bond angles were considered only. 
The CuCC and CCuCu angles (with in-plane bonds) are ignored in the applied model.
Considering these angles requires the specific
optimization of angular parameters which leads to the polarization
of angles that does not fit to the original the AT model.

%%%%%%%%%%%%%%%%%%%%%%%%%%%%%%
% F1
%%%%%%%%%%%%%%%%%%%%%%%%%%%%%
\begin{figure*}[hbtp]
\begin{center}
\includegraphics*[height=6cm,width=8cm,angle=0.]{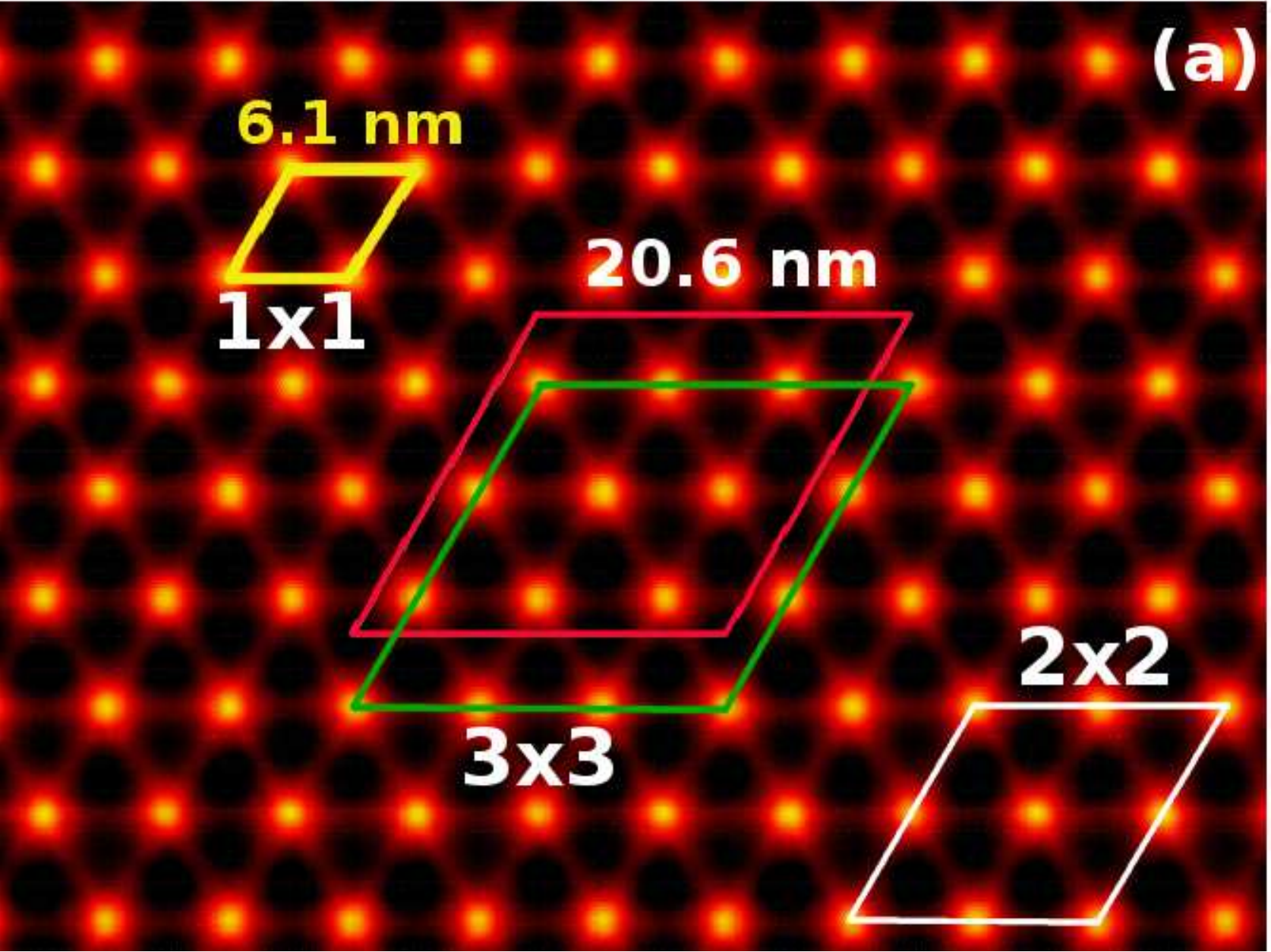}
\includegraphics*[height=6cm,width=8cm,angle=0.]{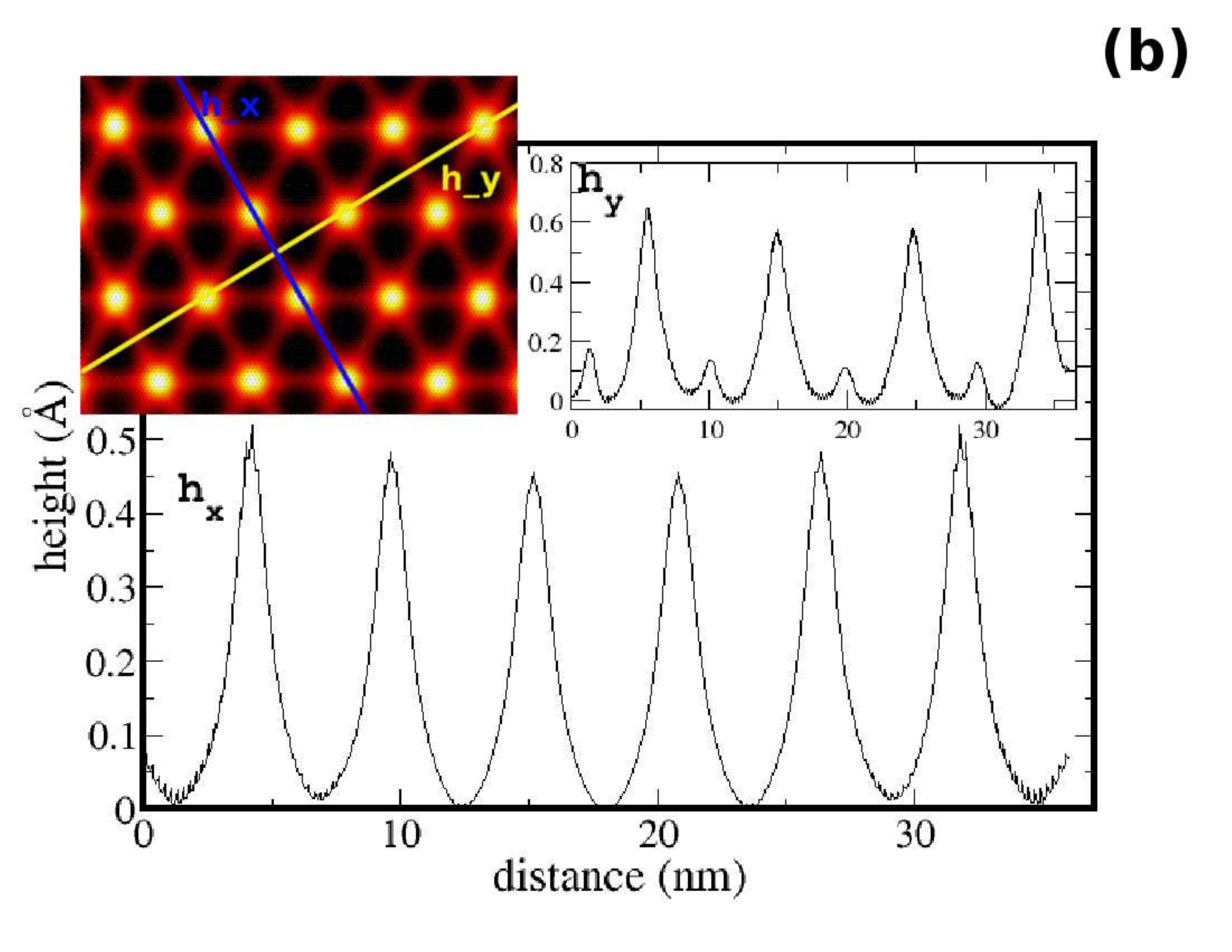}
\includegraphics*[height=6cm,width=8cm,angle=0.]{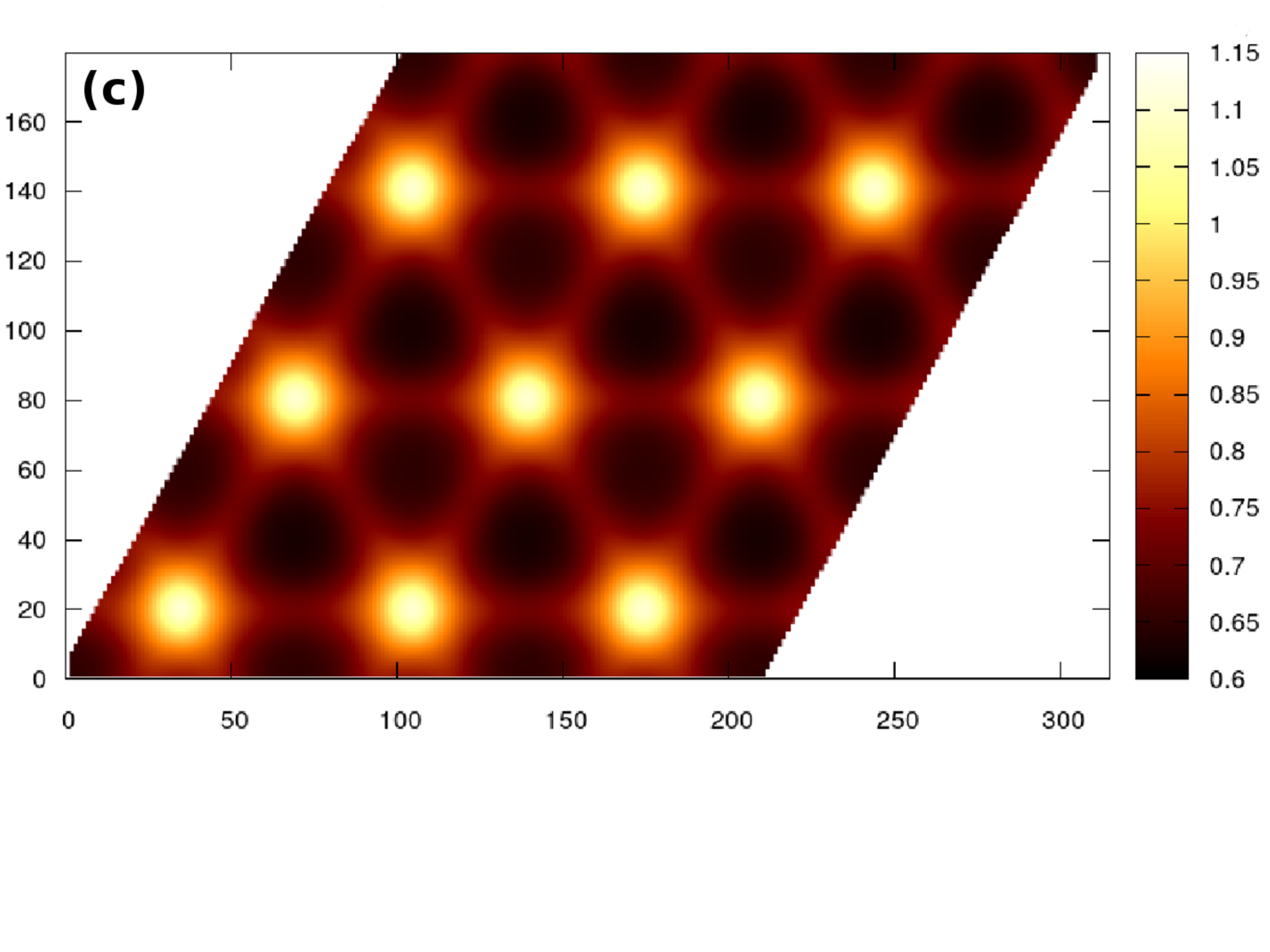}
\includegraphics*[height=6cm,width=8cm,angle=0.]{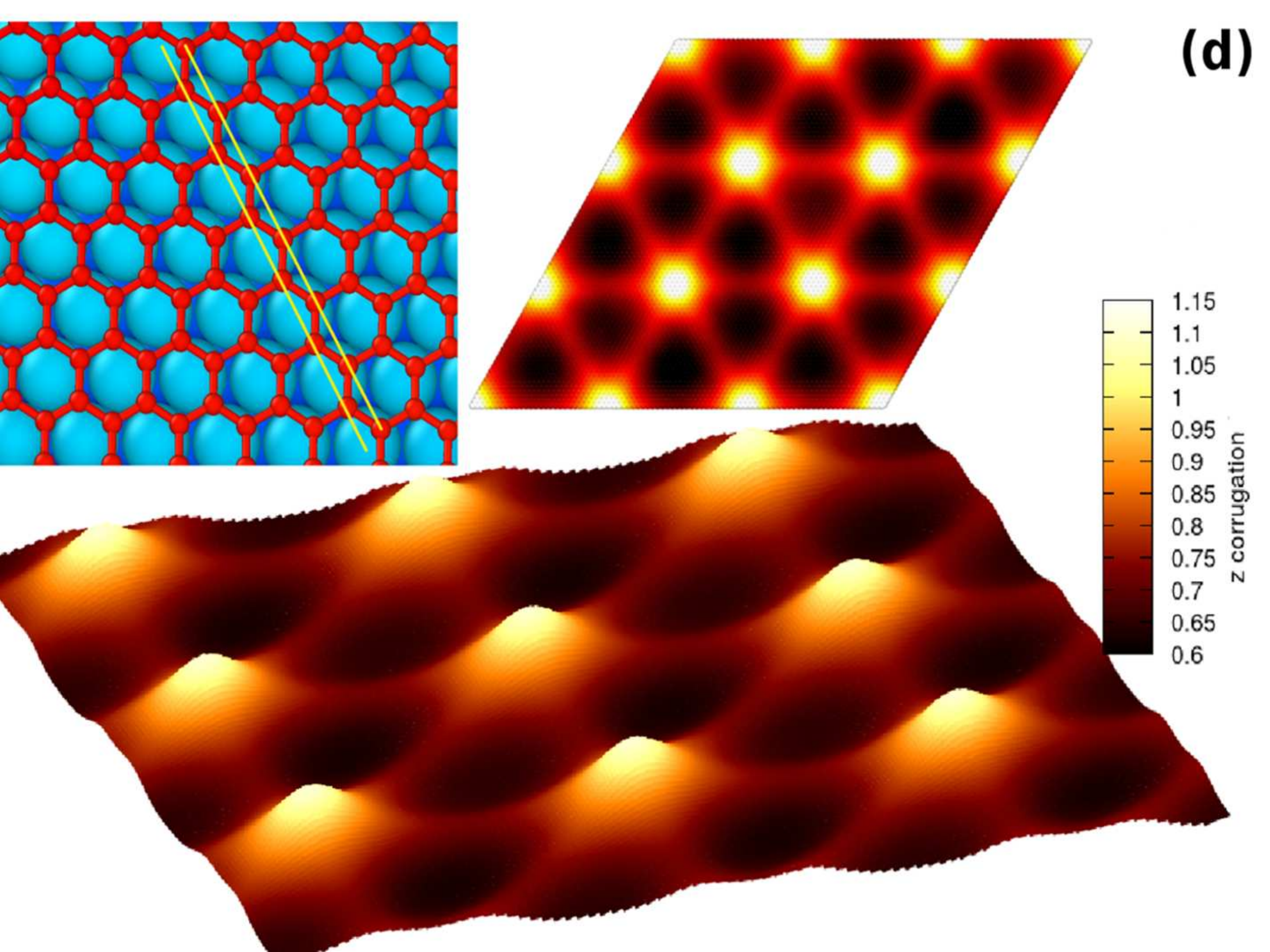}
\caption[]{
 The results of TL-CMD simulations on moir\'{e} superstructures
at 300 K for aligned graphene on Cu(111).
(a) Color coded topographic image. Minimal (yellow, $1 \times 1$) and larger (red, $3 \times 3$ supercells)
are also marked.
In the larger supercell the moir\'{e} protrusions are inequivalent
within a $1 \times 1$ subunit.
The green supercell depicts a nearly equivalent $3 \times 3$ replication array.
(b) Height profiles along the high symmetry directions.
The moir\'{e} superstructure splits into a twofold symmetry pattern.
(c)-(d): Height variations ($\hbox{\AA}$) in the large rhomboid unit supercell of the moire pattern including
14500 Carbon atoms.
Inset on the right in Fig. \ref{F1}(d) shows another equivalent section of the $3 \times 3$
supercell.
Inset on the left in Fig 1(d) depicts the perfect $\Theta = 0.0^{\circ}$
alignment of the gr sheet ($\Theta$ is the misorientation angle of the 
gr sheet with respect to the Cu(111) surface).
The misalignment angle (rotation angle) is calculated
as the angle between the Cu(111) atomic rows on the surface
and between the zig-zag line of the Carbon atoms.
The dimension of the $x$ and $y$ axes are in $\hbox{\AA}$.
%{\em Color coding}: light colors correspond to protrusions (humps) and dark 
%ones to bulged-in regions (bumps).
}
\label{F1}
\end{center}
\end{figure*}
%------------------------------------------------------

%%%%%%%%%%%%%%%%%%%
%\subsection{Simulation rules}
%%%%%%%%%%%%%%%%%%%

\subsection{{\em Ab initio} DFT calculations}

 First principles DFT calculations have also been carried out for
calculating the adhesion energy per Carbon atoms
vs. the C/Cu distance for a small ideal systems with a flat graphene layer.
The obtained potential energy curves (PECs)
can be compared with the similar curve of MD calculations.
We also calculate the DFT potential energy curves of
various binding registries of gr including the hollow and ontop configurations
(atop-fcc and hcp) and also the bridge one.

 For this purpose we used 
the SIESTA code \cite{SIESTA,SIESTA2} which utilizes atomic centered numerical basis set.
The SIESTA code and the implemented Van der Waals functional (denoted as DF2,
LMKLL in the code \cite{SIESTA2}) 
successfully employed in several cases 
for gr (see e.g recent refs. \cite{sies1,sies2}).
We have
used Troullier Martin, norm conserving, relativistic pseudopotentials in fully separable
Kleinman and Bylander form for both carbon and Cu.
A double-$\zeta$ polarization (DZP) basis set was
used.
In particular, 16 valence electrons are considered for Cu atoms
and 4 for C atoms.
Only $\Gamma$ point is used for the k-point grid in the SCF cycle.
The real space grid used to calculate the Hartree, exchange and correlation
contribution to the total energy and Hamiltonian was 300 Ry (Meshcutoff).
% The gradient-corrected Exchange and correlation are calculated
%by the revPBE/DF2 van der Waals functional \cite{Dion}.
%We also employed the local density approximation (LDA) and the
%gradient corrected
%variant PBE \cite{PBE}.
%The Grimme's semiempirical functional \cite{Grimme} has also been used together
%with the PBE/GGA DFT functional \cite{PBE}.
% The system consists of 299 Carbon and 233  Cu atoms (3 layers Cu).  

%%%%%%%%%%%%%%%%%%%%%%%%%%%%%%
% F2 fi2
%%%%%%%%%%%%%%%%%%%%%%%%%%%%%
\begin{figure}[hbtp]
\begin{center}
\includegraphics*[height=6cm,width=8cm,angle=0.]{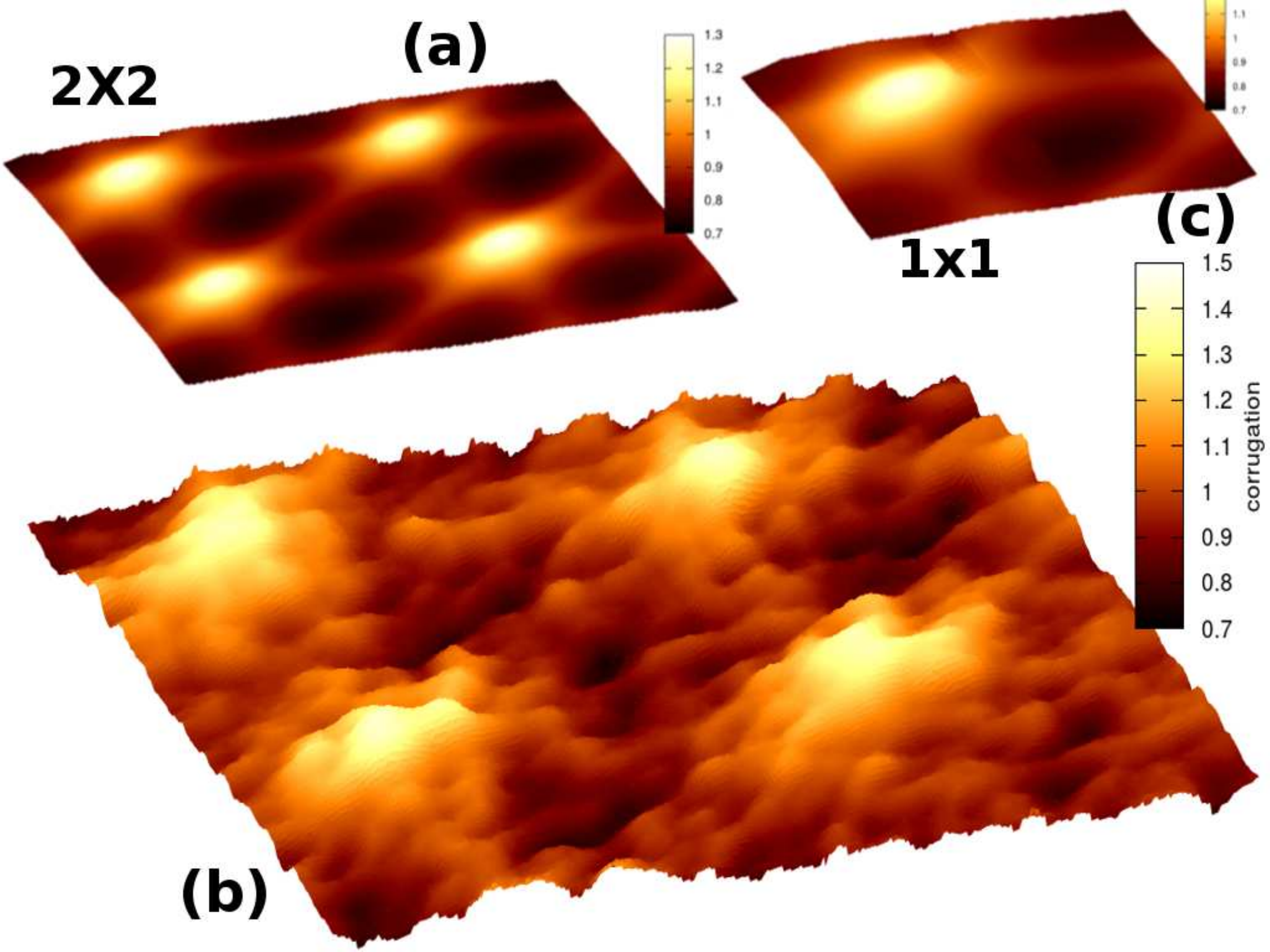}
\caption[]{
Moir\'{e} superlattices as obtained by cg-min/CMD simulations.
The $1 \times 1$ (29 gr honeycombs) and $2 \times 2$ (57 gr honeycombs) 
supercells are shown.
For the $2 \times 2$ superstructure the time averaged ((a), at 10 k time steps) and the
instantaneous (b) images are shown.
(c) The minimal $1 \times 1$ supercell.
}
\label{F2}
\end{center}
\end{figure}
%------------------------------------------------------

%%%%%%%%%%%%%%%%%%%%%%%%%%%%%%
% F3
%%%%%%%%%%%%%%%%%%%%%%%%%%%%%
\begin{figure*}[hbtp]
\begin{center}
\includegraphics*[height=4cm,width=5cm,angle=0.]{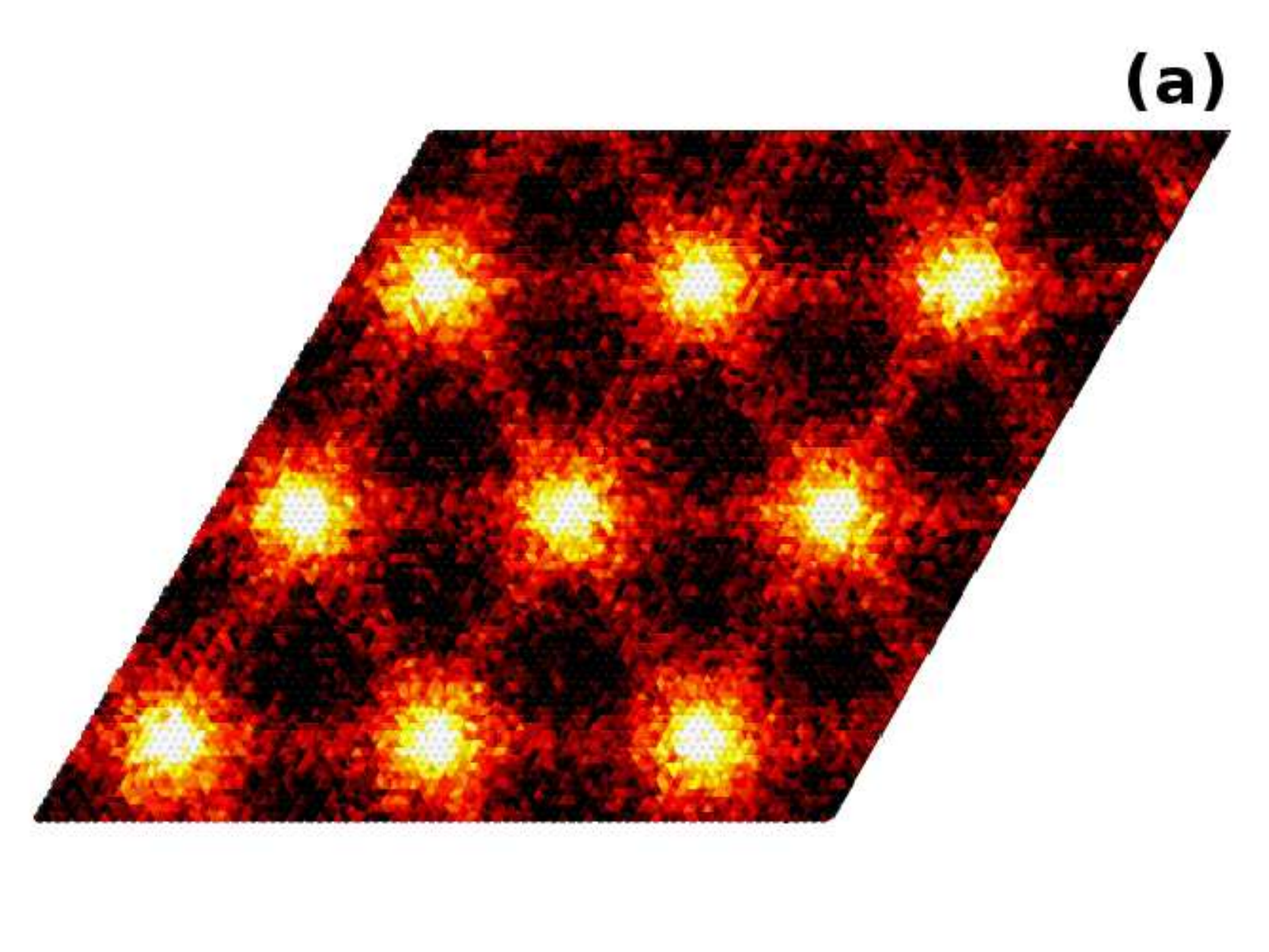}
\includegraphics*[height=4cm,width=5cm,angle=0.]{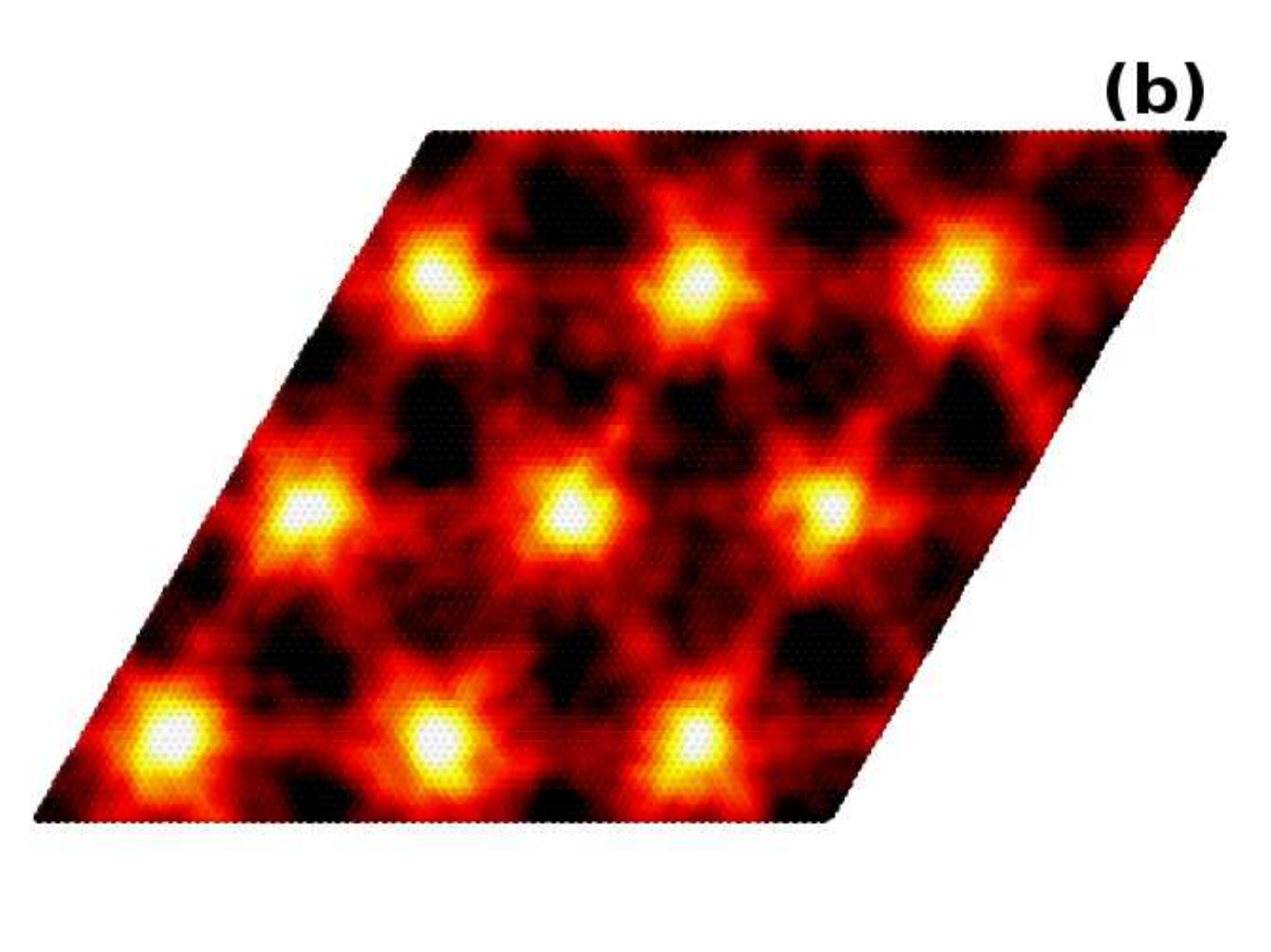}
\includegraphics*[height=4cm,width=5cm,angle=0.]{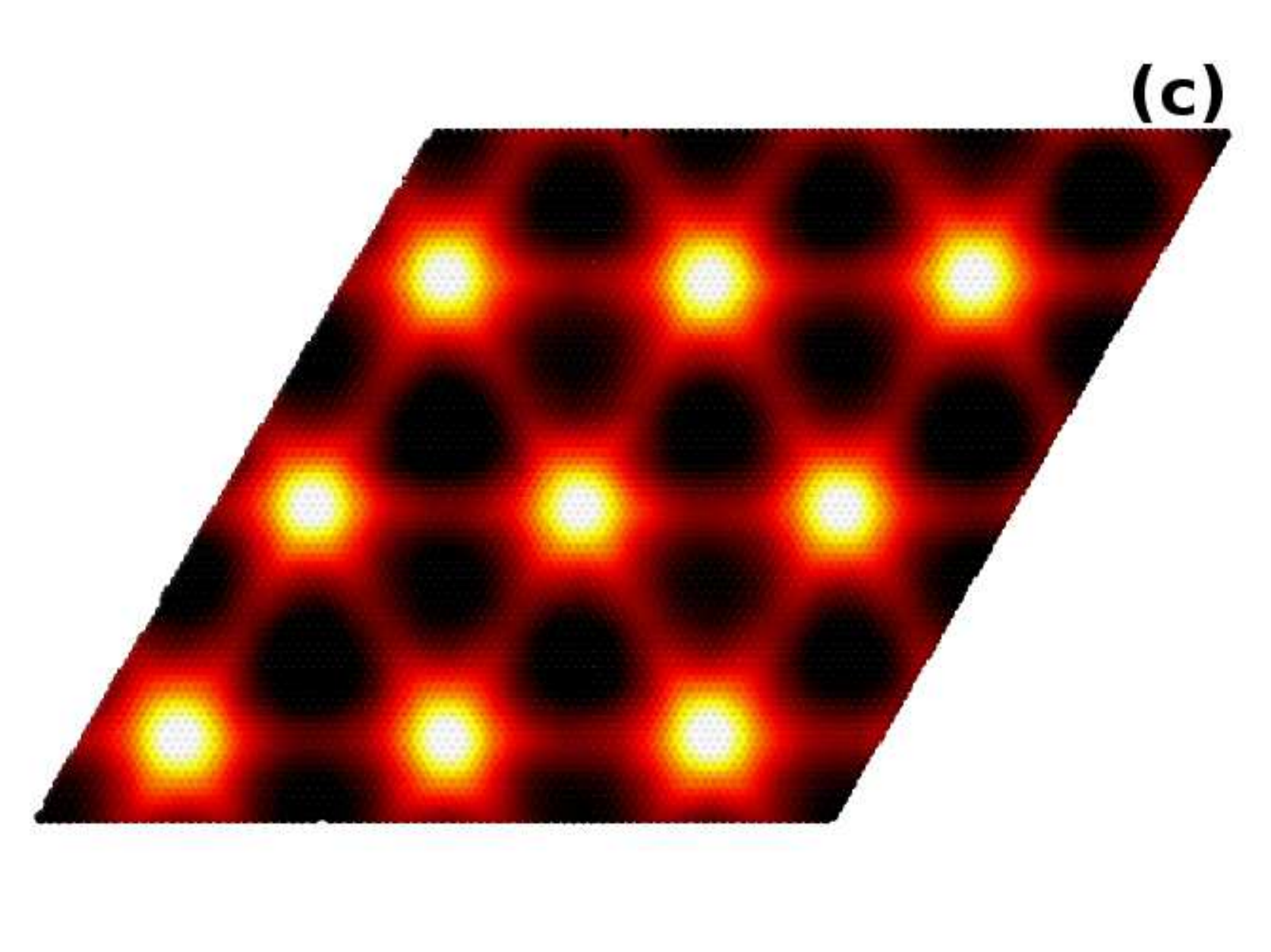}
\caption[]{
 The results of time-lapsed CMD simulations (300 K) time averaged at different time spans
of
100 (a), 1000 (b) and 10000 (c) time steps.
%(d) A not time-averaged (instantaneous) snapshot is shown at 10k steps.
The elapsed time is roughly proportional to the number of steps 
taken place during the simulation
and can be written
as follows: $t_{elapsed} \approx n_{st} \overline{d_t}$,
where $n_{st}$ and $\overline{d_t}$ are the number of time steps and average time step duration
($\overline{d_t} \approx 0.0002$ ps). 
10 k time steps roughly corresponds to the simulation time of $t \approx 2$ ps.
{\em Color coding}: light colors correspond to protrusions (humps) and dark
ones to bulged-in regions (bumps).
}
\label{F3}
\end{center}
\end{figure*}
%------------------------------------------------------

\section{Results and Discussion} 

 With the new parameter set obtained by the parameter fitting procedure
outlined in the Supplementary Material
we were able to simulate gr moir\'{e} superstructures on Cu(111).
The obtained time-lapsed images are shown in Fig. \ref{F1}.
The new AT C-Cu interface potential describes adequately
the weak van der Waals adherence in gr/Cu(111): the adhesion energy
is around 0.14 eV/C which is comparable with the experimental
0.11 eV/C \cite{grcu_adhesion2}.
The average interfacial distance is $3.1$ $\hbox{\AA}$ which is
around the value obtained by a nonlocal vdW-DFT method (present work: 3.05).
Concerning the moir\'{e} superlattice, the main features, such as
the repeat distance of the minimal moir\'{e} cell and
the corrugation are well reproduced (6.1 nm, and 0.055 nm, respectively vs. the experimental 6 nm and $0.035 \pm 0.01$ nm \cite{STM-grcu}). 
Moreover, we found the adequate binding registry: hollow-humps (moir\'{e} hills) and ontop-bumps
(bulged-in regions).
The same binding registry has been identified by DFT calculations
in small gr/Cu(111) systems \cite{Xu_grcu}.
However, the registry of the moirons (moir\'{e} protrusions) has not been studied yet by DFT for larger gr/Cu(111) systems due to the too large size of the supercells (even the minimal cell is too large),
however, it is likely, that the 
optimal position of the moirons is the hollow registry as in gr/Ru(0001).
This is because, by a simple Lennard-Jones (LJ)
pairwise potential for C-Cu interactions, one can get
the incorrect configurations of hollow-bumps and ontop-humps similar to that found for gr/Ru(0001) with LJ potential \cite{Sule13}.
It has been shown recently, that 
LJ provides improper binding sites for gr adherence
and binds Carbon atoms more strongly to the hollow than
to the ontop sites (hcp and fcc sites) \cite{Sule13}.
However, the angular dependent AT C-Cu potential corrects
this deficiency and gives a proper angular orientation at the
gr/Cu(111) interface.
On the basis of these findings mentioned above
we believe that the new force field is suitable for
describing properly vdw-adhesion and the moir\'{e} superlattice in
gr/Cu(111).

%\subsection{Details of parameter fitting}

\subsection{The large corrugated coincidence superlattices} 

% It has been realized after many trial-and-error simulations 
%that the minimal ($1 \times 1$) supercell provides
%unstable periodic structures.
%Increasing the moir\'{e} unit cell the $3 \times 3$ supercell 
%proved to be the optimal choice for this system as
%a stable superlattice.
 In Figs. \ref{F1}(a)-(d) results of CMD simulations with periodic simulation cells are shown for aligned
gr on Cu(111). 
 It turned out after trial simulations that the superlattice shown in Fig. \ref{F1}(a) (red and green rhombuses) and 
Fig. \ref{F1}(c) the $3 \times 3$ supercell is the minimal stable
coincidence superstructure which is nearly commensurate with a 
misfit of 0.64 \%.
The various possible supercells of $N \times N$ contain $N^2$ moirons 
(moir\'{e} hills) or $N^2$ minimal rhombuses as shown for
the $3 \times 3$ supercell in Fig. 1(a).
In the $1 \times 1$ and $2 \times 2$ supercells the lattice mismatch is
1.72 \% and 0.04 \% respectively.
In the much larger $7 \times 7$ coincidence cell the misfit is still
0.24 \%.

 In the $1 \times 1$, $2 \times 2$ and $3 \times 3$ superstructures
$29 \times 29$, $57 \times 57$ and $85 \times 85$ gr honeycombs sit
on the substrate, respectively, nearly commensurately with the support's lattice.
The gr superstructures are commensurate with 27, 55 and 83 Cu(111) atoms, respectively. 
According to the lattice mismatch values the
$2 \times 2$ periodic cells provides the most stable
minimal moir\'{e} superlattice.
The $1 \times 1$ supercell contains only a single moiron and 1682 Carbon atoms.
(see Fig. \ref{F2}(c)).
The $2 \times 2$ (Figs. 2(a)-(b)) and $3 \times 3$ supercells include
6498 and 14450 Carbon atoms, respectively.

 These findings are in line with the results obtained by 
Iannuzzi {\em et al.} recently
for gr/Ru(0001) using vdW-DFT method with geometry optimization \cite{DFT:Ru-Hutter}.
In their work they have demonstrated that 
the unit cell of the moir\'{e} superstructure is the
larger $2 \times 2$ coincidence cell in which the moir\'{e}
hills are inequivalent \cite{grru13}.
We argue that in the case of gr/Cu(111) although the $2 \times 2$ superlattice
could also be sufficient, however, we find the larger
$3 \times 3$ supercell is more adequate for the description of
the periodic moir\'{e} pattern.
%This is the reminiscent of the additional long range order expressed by
%the surface moir\'{e} lattice \cite{Hermann}.

 In order to further support the superior stability one of the possible superlattices
the time averaged cohesive energy of Carbon atoms has been calculated vs.
$N$ (the number of unit cells in the reputation array $N \times N$).
The following values has been found by CMD simulations: -7.460, -7.464, -7.466, -7.460 and -7.461
eV/atom for N=1,2,3,4 and 5.
The minimal supercell of $1 \times 1$ is very close in energetic stability
to much larger systems 
%The $1 \times 1$ supercell shows also a relatively high level stability,
although in this case the moirons are forced to be equivalent.
%There is no possibility of additional freedoms of relaxation 
%and the minimal supercell remains mismatched.
The relative energetics of $N \times N$ supercells
does not convincingly support the superior stability of any of the superlattices.
Nevertheless we use the larger coincidence supercell of $3 \times 3$
for further analysis which is capable of involving the 
key ingredients of superlattice dynamics.

\subsection{Time lapsed moir\'{e} superlattice}

\begin{figure*}[hbtp]
\begin{center}
\includegraphics*[height=5cm,width=7cm,angle=0.]{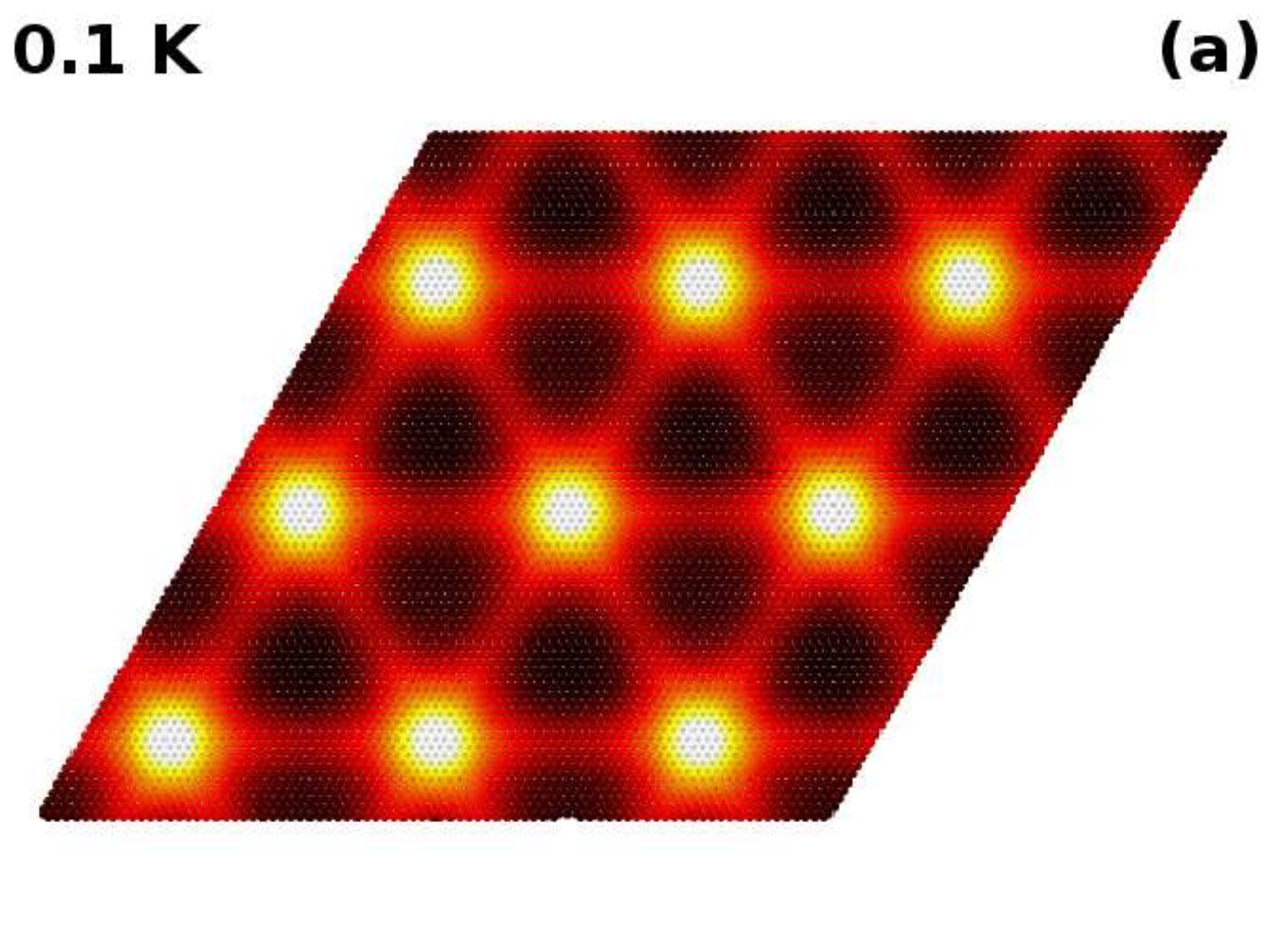}
\includegraphics*[height=5cm,width=7cm,angle=0.]{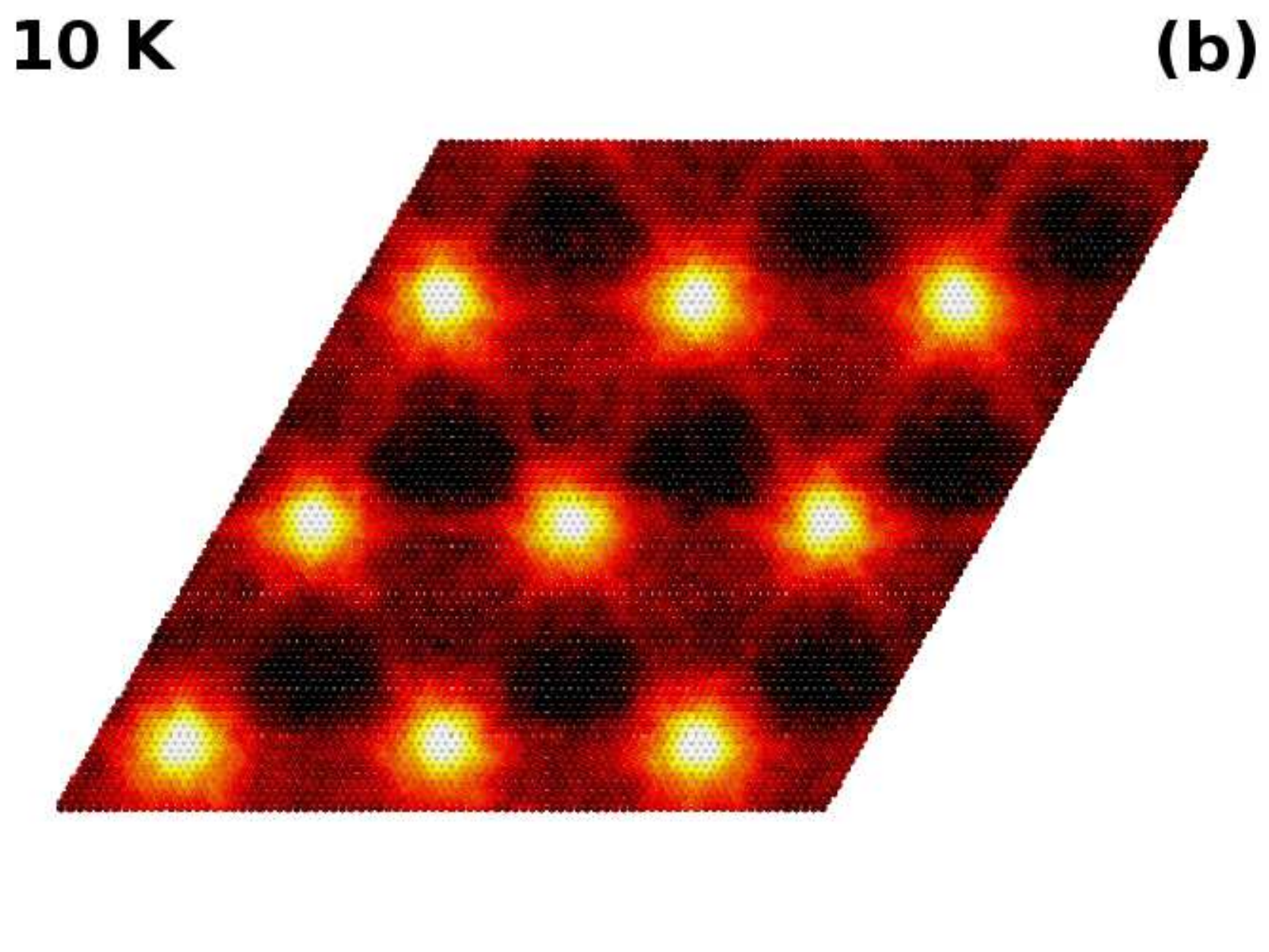}
\includegraphics*[height=5cm,width=7cm,angle=0.]{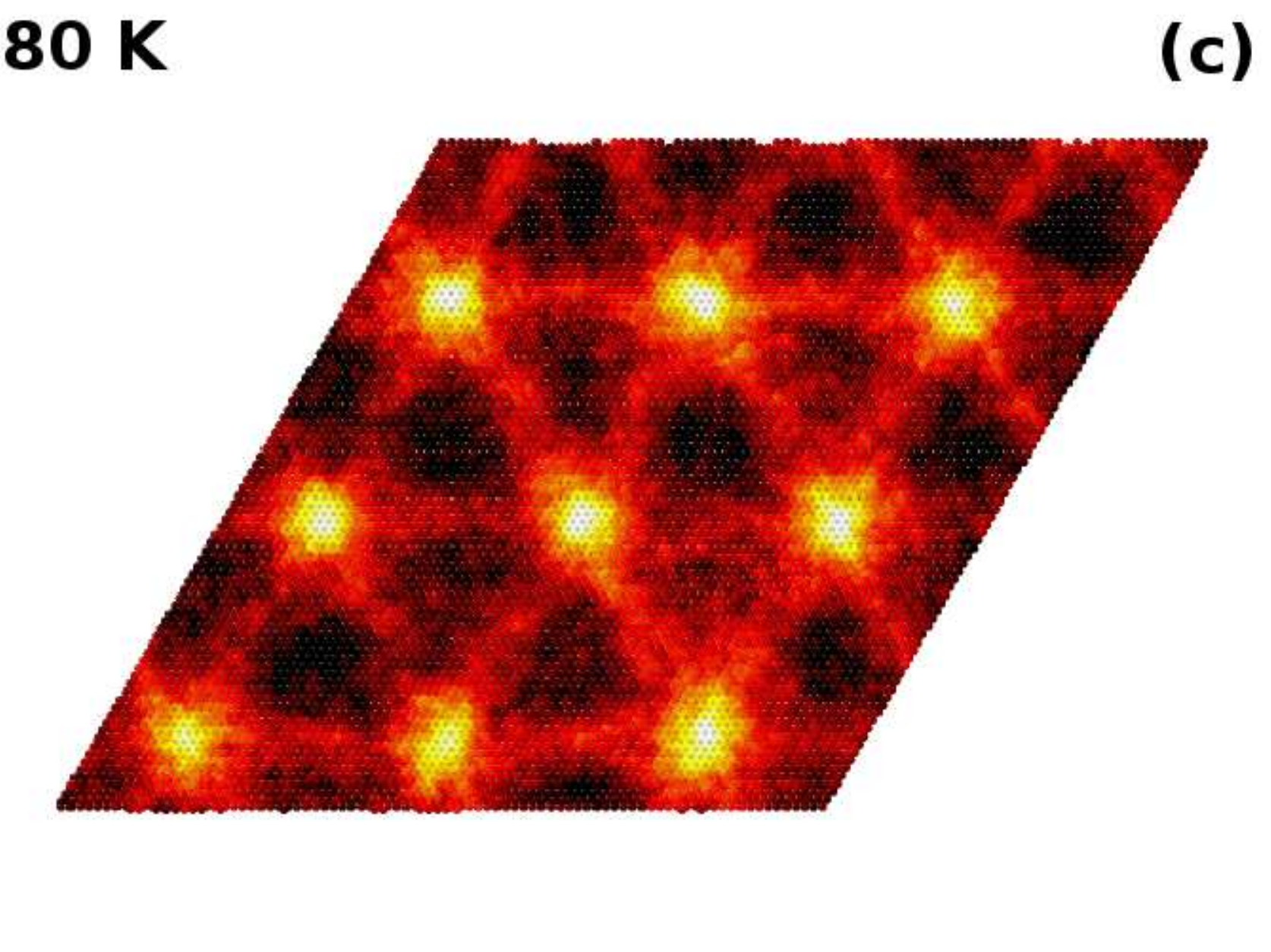}
\includegraphics*[height=5cm,width=7cm,angle=0.]{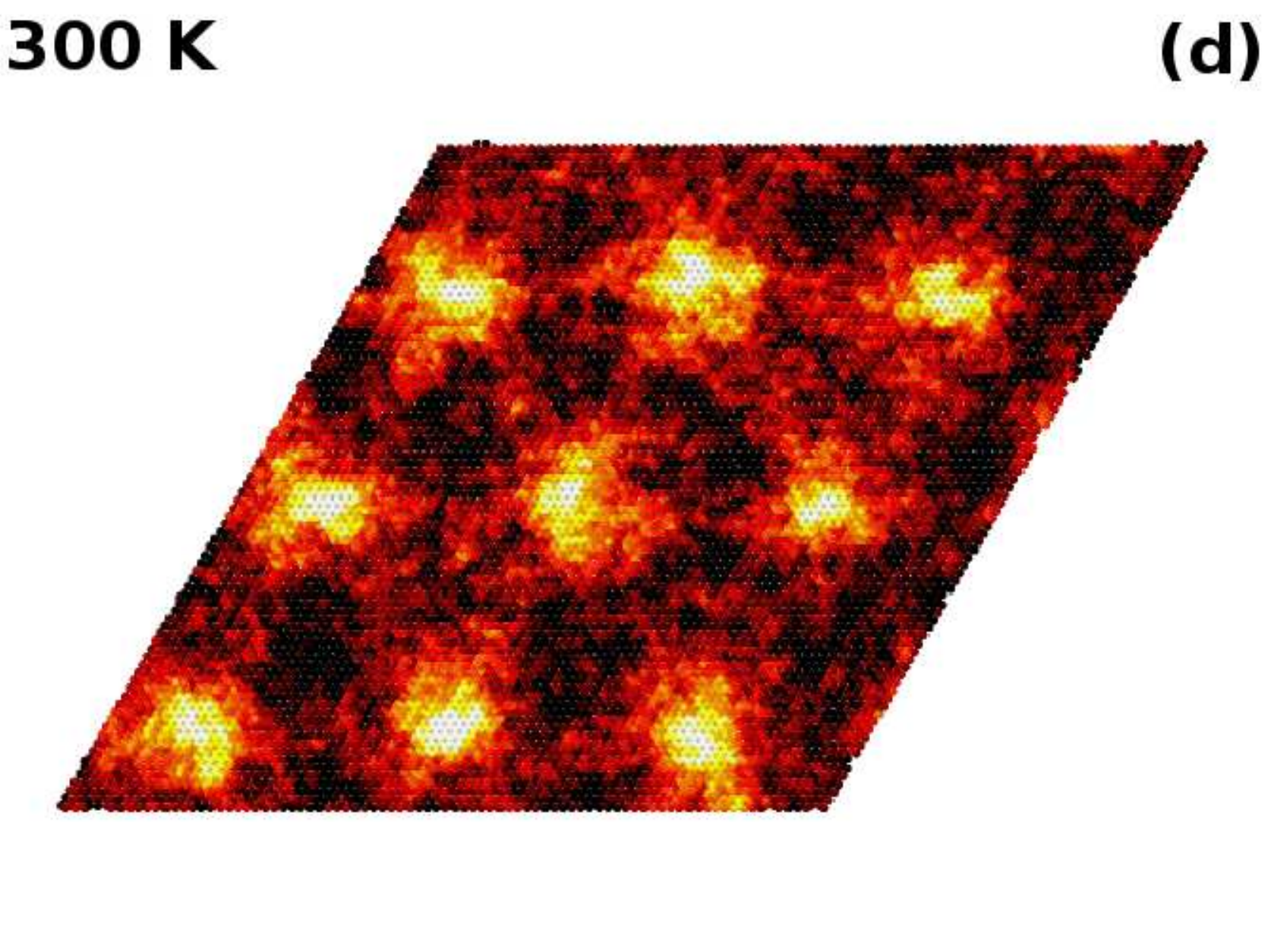}
\caption[]{
 The results of CMD simulations at different temperatures
of
%10 (a), 80 (b), 300 (c) and 500 K.
0.1 (a) 10 (b), 80 (c) and 300 (d) K.
Note, that these typical images were taken from 
instantaneous
snapshots which were extracted at 10 k simulation steps.
{\em Color coding}: light colors correspond to protrusions (moir\'{e} humps) and dark
ones to bulged-in regions (bumps).

}
\label{F4}
\end{center}
\end{figure*}

 In Figs. \ref{F3} the effect of time lapsing on the moir\'{e} pattern is shown.
In Figs. \ref{F3}(a)-(c) the time averaged images are shown with various time span.
A sharp pattern is obtained at or above 10k time steps.
The pattern is never sharp on
instantaneous snapshots such as shown in Figs \ref{F4}(d).
This implies that the experimentally seen images are also in fact time lapsed
patterns. 
In particular, one can see the partial disruption of the pattern:
e.g. the weak "dissolution" of the moirons in their neighborhood.
However, this process if time averaged on a ps time scale, a sharp
moir\'{e} pattern appears.
The underlying consequence of this mechanism could be that
the gr sheet dynamically exhibits an unexpected anisotropy:
the surface reconstruction leads to a disordered pattern and becomes ordered
beyond a certain time interval only.
%If it so than the anisotropic moir\'{e} superlattice should also influence the electronic structure of gr, e.g. leading to anisotropic Dirac cones \cite{Rusponi10}.
The weak disordering of the moir\'{e} superlattice is a dynamical phenomenon
and works on a ps time scale and leads to the break down of the hexagonal
symmetry of the moir\'{e} superstructures.

%%%%%%%%%%%%%%%%%%%%%%%%%%%%%%
% F4 Fig_4
%%%%%%%%%%%%%%%%%%%%%%%%%%%%%
%------------------------------------------------------

 Contrary to the stronger adhesion in gr/Ru(0001)
dynamic moir\'{e} pattern has also been found (not shown here), although it has not
fully been realized recently in ref. \cite{Sule13} that
the time averaged pattern is the real moir\'{e} pattern.
This is because in gr/Ru(0001 the instantaneous images are much closer
to the time lapsed pattern due to the stronger adherence.
In gr/Cu(111), the weaker van der Waals interaction at the interface
results in in some sense free-standing behavior which favors
disordering.
However, contrary to the weak adhesion, moir\'{e} order is partly
retained and the unset of disorder is still under controll.
The weak adsorption energy of gr, however, permits the thermal out-of-plane fluctuation
of the sheet.

 The local variations of the moir\'{e} structure has been found recently
by AFM and LEED in gr/Ir(111) \cite{dynamic_LEED}.
In particular, it has been found that the measured corrugation varies smoothly over several 
moir\'{e} unit cells \cite{dynamic_LEED}.
This could be due to the dynamic moir\'{e} structure mechanism outlined
above. In our time averaged structures we also find the slight variation
of the bump-to-hump corrugation with some 10 pm which is comparable with
that found in ref. \cite{dynamic_LEED} (12 pm).
In these calculations a larger area (rhombic supercell: $255 \times 255$ unit cell$^2$ of the honeycomb lattice) is sampled.
The spatial variation of corrugation occurs not only at certain time points
but also time-to-time the height of a given moiron varies slightly.
This is in line with the conclusion of ref. \cite{dynamic_LEED} that
there is a 2nd order moir\'{e} which is not expected to be rigid 
and likely to exhibit fluctuations.
The 2nd order moir\'{e} is due to the weak oscillation of occupied
positions by Carbon atoms around the idealistic 1st order moir\'{e} structure.
The latter one is seen by time averaged CMD simulations and the former one
in instantaneous snapshots such as shown in Fig. \ref{F3}(c).

 {\em Therefore we argue that the coexistence of competing moir\'{e} orders
forms the final observable pattern seen by STM}.
This is surprising since it has been widely accepted that moir\'{e} superstructures
are nearly static objects due to the fact that the moir\'{e} image emerges from the superposition
of the relatively rigid graphene lattice on the support's lattice.
While the assumption of lattice rigidity is more or less holds,
however, the temperature induced out-of-plane vibrations and the relative lateral mobility and/or the rotation of
the lattices are not necessarily negligible.

 The temperature dependence of the dynamical distortion of the moir\'{e} pattern
has been studied in detail and typical instantaneous simulated images
are shown on Figs. \ref{F3}(a)-(d) as obtained at 10 k time steps.
Close to zero K the pattern remains still nearly sharp, however,
with increasing temperature disorder sets in more and more
strongly. The time averaged images show nearly perfect moir\'{e} order.

%%%%%%%%%%%%%%%%%%%%%%%%%%%%%%
% F5 Fig_5
%%%%%%%%%%%%%%%%%%%%%%%%%%%%%
\begin{figure*}[hbtp]
\begin{center}
\includegraphics*[height=10cm,width=15cm]{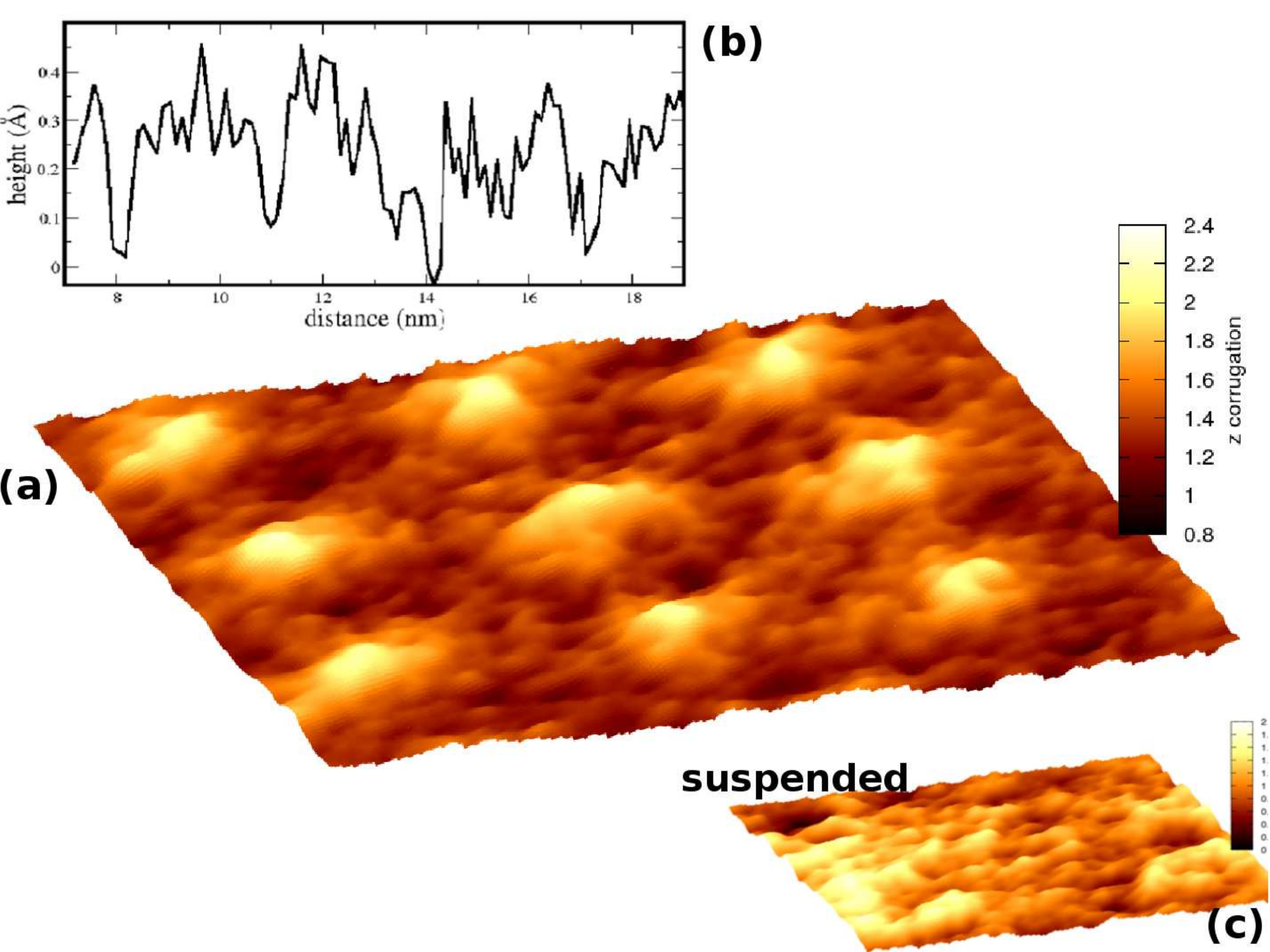}
\caption[]{
The instantaneous snapshot of the topography of the gr/Cu(111) moir\'{e} superlattice
at 300 K and at 5 k time steps (a).
Corrugation height profile is also shown.
The corrugation are in $\hbox{\AA}$. 
%Note the surprisingly deep local regions where the corrugation
%can exceed 0.15 nm.
Inset (b): a typical height profile along horizontal direction in the middle of the system.
(c) For comparison the rippled suspended (freestanding) graphene system is also shown at 300 K.
The height variation extends in the range of $[0;2]$ $\hbox{\AA}$.
Periodic boundary conditions were also maintained in suspended
gr simulations in order to ensure similar conditions as in 
supported gr simulations.
{\em Color coding}: light and dark colors correspond to protrusions and 
to bulged-in regions, respectively.
}
\label{F5}
\end{center}
\end{figure*}
%------------------------------------------------------

 In Fig \ref{F5} a more refined instantaneous image is shown for 300 K as obtained at 10 k time steps. 
The image reveals that corrugation is unexpectedly different from 
the time lapsed images shown on Fig. \ref{F3}(c) and Figs. \ref{F1}(c)-(d).
The amplitude of height variation can exceed locally the 0.15 nm which
is time averaged to the much lower 0.05 nm.
The topography at 300 K becomes then rather wrinkled which leads locally to
bulged-in regions.
The hexagonal shape of the time lapsed moir\'{e} protrusions becomes
irregular.
In Inset Fig. 5(b) the height profile along a thin section of the surface
also reveals the irregular variation of the height profile 
which is in contrast with the time lapsed regular profile shown in Fig. \ref{F1}(b).

 These findings are somewhat surprising since the generally accepted regularity of supported gr 
is challenged now. 
The regularly rippeled surface of gr might be a time averaged phenomenon. 
Even ultraflat gr \cite{gr_ultraflat} is in fact can be a dynamical system with a relatively high out-of-plane
amplitudes.
Therefore, gr within ultra-short time scales (ps and less)
shows a much different topography than the available microscopy images.
In particular, local height fluctuations partly destroy the first order moir\'{e} pattern 
seen on time lapsed images.
The competition of height fluctuation and the moir\'{e} pattern might
influence the various properties of gr such as the gap, band structure
and transport properties.
The interplay between lattice deformations and electron
dynamics is an important ingredient to understand
and control the electronic properties of graphene
devices.
%%%%%%%%%%%%%%%%%%%%%%%%
% T1
%%%%%%%%%%%%%%%%%%%%%%%%
%\LARGE{
\begin{table*}[tb]
%\begin{table}[t]
%\raggedleft
\raggedright
%\sidecaption
\caption[]
{
The summary of various properties obtained for gr/Cu(111)
by classical molecular dynamics simulations using the fitted Abell-Tersoff
potential for the interface.
The main properties of the moir\'{e} superstructures.
}
\label{T1}
\begin{ruledtabular}
\begin{tabular}{llcccccccc}
  & &  &  &  &  &  &  &  &   \\
   method & $d_{per}$ (nm) & $\xi$ ($\hbox{\AA}$)  & $\xi_{Cu}$ ($\hbox{\AA}$) & $d_{ave}$ ($\hbox{\AA}$) & $a_{gr}$ ($\hbox{\AA}$) & $a_{lm}$ ($\%$) & $E_{adh}$ (eV/C) & $\Delta E$  & $E_{gr}$  \\
  & &  &   &  &  &  &  &  &   \\
\hline
  & &  &   &  &  &  &  &  &   \\
  {\em freestanding}  & &  &   &  &  &  &  &  &   \\
  & &  &   &  &  &  &  &  &   \\
     &  susp-min & 2.6 & - & - & 2.462 & - & - & 0.0 & -7.408 \\
     &  susp-TL/CMD & 2.0 & - & - & 2.462 & - & - & -0.013 & -7.383 \\
     &  flat-min & 0.0 & - & - &  2.462 & - & - & 0.0 & -7.408 \\
     &  flat-TL/CMD & 0.0 & - & - &  2.459 & - & - & 0.0 & -7.370 \\
  & &  &   &  &  &  &  &  &   \\

  {\em supported} & &  &   &  &  &  &  &  &   \\
  & &  &   &  &  &  &  &  &   \\
 MIN &  6.1 & $0.51$ & $0.22$ & 2.98 & 2.459 & 3.56 & -0.145 & 0.001 & -7.407 \\
 CMD &  6.1 & $0.55$ & $0.26$ & 2.99 & 2.457 & 3.56 & -0.146 & -0.016 & -7.386 \\
%  &  &  &  &  &  &  &  &  &   \\

  & &  &  &  &  &  &  &  &   \\
 EXP & 6.0$^a$ & $0.35 \pm 0.1^a$ &  n/a &  n/a & 2.46 & 3.53 & -0.11$^b$ & n/a & n/a \\

  & &  &  &  &  &  &  &  &   \\
 DFT & n/a & n/a & n/a &  3.25$^c$, 3.05$^d$ & n/a & n/a & -0.062$^c$, -0.198$^d$ & n/a & n/a \\
  &  &  &  &  &  &  &  &  &   \\
\end{tabular}
%\end{ruledtabular}
%\footnotemark*{
\footnotetext*[1]{
pw denotes present work,
%$\Theta$ is the rotation angle in degree (the angle between
%the line of zig-zag Carbon atoms and the adjacent Cu(111) atoms along the line (110) shown in Inset Fig. 2(a).)
$d_{per}$ is the periodicity of the minimal moir\'{e} pattern
(the edge length of the rhombus with 4 moir\'{e} humps, $1 \times 1$ supercell),
$\xi$ and $\xi_{Cu}$ are the average corrugation for gr and the topmost
Cu(111) layer ($\hbox{\AA}$).
$d_{ave}$ is the average inter-layer (C-Cu) distance ($\hbox{\AA}$) at the interface.
$a_{gr}$, $a_{lm}$ are the lattice constant of gr ($\hbox{\AA}$) and
the lattice mismatch ($\%$) after simulations ($a_{lm}=100 (a_{s}-a_{gr})/a_{gr}$).
\underline{CMD}: pw, fitted Abell-Tersoff results with cg minimization with CMD at 300 K (gr/Cu(111)).
{\em graphene-only simulations}:
\underline{{\em susp}}: periodic suspended (freestanding) graphene simulations which lead essentially
to randomly rippled gr (300 K),
\underline{{\em flat}}: periodic 2d CMD simulations with a freestanding flat graphene (300 K),
\underline{{\em min}}: periodic cg-minimization only (geometry optimization together with box relaxation),
\underline{EXP}: the experimental results:
corrugation ($\xi$): our STM results, 
\underline{DFT} results are also given for comparison \cite{DFT:Ru-Hutter,Batzill,DFT:Ru_Wang}.
%DFT (pw): PBE \cite{PBE} and revPBE/vdW-DF2 \cite{Dion} results as obtained by the author. See further details
%in the caption of Table I.
All quantities are given per Carbon atom.
The adhesion energy $E_{adh}=E_{tot}-E_{no12}$, where $E_{tot}$ is the potential energy/C
after md simulation. $E_{no12}$ can be calculated using the final
geometry of md simulation with heteronuclear interactions
switched off. Therefore, $E_{adh}$ contains only contributions from
interfacial interactions.
%$E_{str,gr}$, the strain energy of the corrugated gr-sheet,
$\Delta E$ (eV/C) is the energy difference with respect to the
perfectly flat periodic gr.
$\Delta E =E_{gr}-E_{gr,flat}$,
where $E_{gr}$ and $E_{gr,flat}=-7.37$ eV/C are the cohesive energy of C atoms
%where $E_{gr}$ and $E_{gr,flat}=-7.426$ eV/C are the cohesive energy of C atoms
in the corrugated and in the relaxed periodic flat (reference) gr sheet as obtained
by the AIREBO C-potential \cite{airebo}.
%#######################################
%# xxl system (C-only), ~500k atom : -7.4180893 eV/C
%#######################################
$^a$ from refs. \cite{Gao,STM-grcu},
$^b$ from ref. \cite{grcu_adhesion2}., double cantilever beam method:
$E_{adh}$=0.72 J/m$^2$.
%$^c$ from ref. \cite{grcu_adhesion2},
%$^d$ from ref. \cite{JACS-grcu},
%Another technique (nano-scratch technique) gives an unexpectedly high value: 12.8 J/m$^2$ \cite{grcu_adhesion}.
%$^e$ present STM work,
$^c$ from ref. \cite{RPA}, obtained by accurate random phase approximation
for a very small modell system.
$^d$ present work: nonlocal vdw-DFT calculation for a small flat system (463 atoms):
hcp: -0.198 eV/C ($d_0=2.95$ $\hbox{\AA}$), hollow: -0.182 eV/C ($d_0=3.09$ $\hbox{\AA}$), vdw-DFT geometry optimized
structures: ontop hcp: -0.350 eV/C, hollow: -0.133 eV/C.
}
\end{ruledtabular}
\end{table*}
%}
%%%%%%%%%%%%%%%%%%%%%%%%%%%%%%

 The amplification of structural fluctuations in suspended gr
has been known for a while
called intrinsic corrugation (ripples) \cite{Meyer,Fasolino}, however, little is known about similar
features in supported gr.
The topographic image of a typical rippled suspended gr system (gr-Cu(111) interaction is switched off) is shown in Fig. \ref{F5}(c).
The ripples are randomly distributed here as opposed to Fig. \ref{F5}(a)
in which ripples are still ordered embedded in the background "noise"
of the randomly arranged height fluctuations.
The latter one are similar to that seen on Fig. \ref{F5}(c).

 The fluctuations of the
corrugation, called flexural phonons, have been proposed
to be the source of the intrinsic limit in the electronic
mobility of graphene suspended samples \cite{flexph}.
It has also been found recently that intrinsic corrugation
is partly damped in supported gr
although not fully suppressed \cite{rippledgr_sio2,extrinsic}. 
Therefore the observed random height fluctuation (can be seen in Fig. \ref{F5}(b))
can be attributed to the persistence of intrinsic corrugation
which competes with moir\'{e} ordering induced by the partial conformation of the gr sheet to the substrate.

 Therefore contrary to the presence of the moir\'{e} order
as a time averaged pattern
 {\em the relatively high amplitude height variations
could significantly influence the performance of electronic
devices made from supported gr.}
However,
higher adhesion leads to smaller intrinsic noise
as can be seen in gr/Ru(0001) \cite{Sule13}.

 It should also be noted that the observed height fluctuations are different
from other reported buckling such as the one e.g. on the elastic response of
gr nanodomes \cite{buckling}.
While the periodic buckling of the moir\'{e} humps 
is present and can be induced or amplified by the AFM tip in contact mode
with decreasing tip-sample distance \cite{buckling}, 
the random height fluctuation of the periodic moir\'{e} pattern presented
in this work as already mentioned above is intrinsic.
Unfortunately, these random intrinsic height fluctuations of gr
are not visible by commonly used experimental methods
because
ultrafast nanoscale processes are well beyond the spatial and temporal resolution limits
of current scanning probe characterization techniques.
Time-resolved surface X-ray diffraction \cite{timeres_SXRD} could offer
a way in the near future to probe the ps-scale dynamics of the moir\'{e} superlattice.

%Unfortunately, atomic-scale investigations of dislocation evolution in a bulk object are well beyond the spatial and temporal resolution limits of current characterization techniques. Here we overcome the experimental limits by investigating the two-dimensional graphene in an 

%When the dimensionality is reduced, height fluctuations
%are amplified due to the known tendency to in-
%stabilities in low dimensions. 

\subsection{Details of structural and energetic properties} 

 In Table I. the various structural and energetic properties of the simulated rotated structures
have been summarized.
The notable features are the following:

 In spite of the significant lattice misfit of $a_{lm} < 3.56$ $\%$
 the aligned moir\'{e} corrugated gr phase is slightly deeper in energy than
the perfectly relaxed flat gr, the energy difference
$\Delta E \approx -0.016$ eV/C).
% significant ($\xi \approx 0.45$ $\hbox{\AA}$).
This can be attributed to the efficient strain relief in the large coincidence supercells \cite{grru13}.
The cohesive energy of the suspended (freestanding) rippled gr shown in Fig. 5(c) is
-7.383 eV/C which is very similar to that of the periodically moir\'{e} patterned supported gr
(-7.386 eV/C).
Therefore, gr in its bound state is as stable as the rippled sheet
which is intrinsically buckled as obtained by the CMD simulations of the present work.
The $\Delta E \approx 0.016$ eV/C is, however, below the magnitude
of thermal motion at 300 K (0.026 eV).
%Even the rippled gr is more stable than the flat one which is due to 
%intrinsic rippling \cite{Meyer}.
%Nevertheless, the conformation of gr into substrates leads to energy gain
%with respect to the freestanding form even without the energy gain
%provided by adhesion.

%ovided by adhesion.We found $0.16$ \% and $0.04$ \% mismatch in the minimal and large supercells,
%respectively.

% Concerning the energetic stability with respect to the perfectly
%flat periodic gr,
% $\Delta E \approx 0.04$ eV/C (see Table I.,) which is somewhat above
%the magnitude of thermal motion ($\sim 0.026$ eV/K at 300 K)
%which in principle renders this phase detectable even at room temperature.

% The {\em adhesion energy} has also been calculated and is shown
%in Table I.
%We find a somewhat larger adsorption energy of $-0.145$ eV/C
%than by experiment ($-0.11$ eV/C, \cite{grcu_adhesion2}).
%In general, the binding energy of gr to Cu(111) is smaller than
%in other systems, such as gr/Ru(0001) ($0.17-0.2$ eV/C, see e.g. ref. \cite{Sule13}) and larger
%than in the weakest bound systems (e.g. gr/SiO$_2$, $E_{adh} \approx -0.07$ eV/C) \cite{grsio2}.

\section{Conclusions}

 A new C-Cu interfacial force field has been developed for the graphene/Cu(111)
system with which we are able to describe adequately moir\'{e}
superlattice formation.
We find that a large $3 \times 3$ coincidence unit cell (the unit cell of the superlattice)
reproduces many important properties of the system such as corrugation, missorientations, moir\'{e} superlattice
and adhesion energy.
A stable and sharp moir\'{e} pattern becomes visible by time-lapsing over at least 10 k simulation steps.
The instantaneous images shows up disordered pattern in various magnitude
depending on the temperature.
The competition between the intrinsic corrugation induced disordering and moir\'{e} ordering
leads finally to regular moir\'{e} superlattice in time averaged images
in a ps timescale.
The competing moir\'{e} order is driven by inhomogeneous local out-of-plane fluctuations (disordered buckling instability).

 The present study reveals then that supported gr
is much more rippled than is widely accepted.
The constant presence of the large amplitude (0.15 nm)
height fluctuations might deteriorate the 
performance of supported gr as an electronic device.
The electrons encounter and interact with the out-of-plane vibrations of the gr-sheet, and that can affect the material’s conductivity.
Stronger adhesion could help, though the minimization of the amplitude
of the height variations to the lowest possible level.

%
% 1 degree kelvin= 8.621738 X10-5 eV
% 298 K = 0.025693 eV
%

\section{Supplementary Material}
Supplementary Material is available on parameter fitting
and on the details of the force field.

\section{acknowledgement}
%{\scriptsize
%This work is supported by the OTKA grant
% K-68312
%from the Hungarian Academy of Sciences.
%Support from the bilateral German-Hungarian
%exchange program DAAD-M\"OB (Grant No. 37-3/2008)
%and German Science Foundation (DFG research
%group 845, project HE2137/4-1) is also
%acknowledged.
The calculations (simulations) have been
done mostly on the supercomputers
of the NIIF center (Hungary).
%The kind help of P. Erhart (Darmstadt) in the usage of the PONTIFIX code is greatly acknowledged.
The availability of codes LAMMPS (S. Plimpton) and OVITO (A. Stukowski) are also greatly acknowledged.

%\appendix
%Supplementary Material is available on parameter fitting.

%\vspace{-0.7cm}

%\end{document}

\section{Supplementary Material} 
\vspace{1cm}
\bec
%{\em Time-lapsed graphene moir\'{e} superlattice on Cu(111)\\
{\em The details of parameter fitting for the gr/Cu(111) interface
}
\ec

%\begin{abstract}
 The force field parameters were fitted against a data set including few small representatives of gr/Cu(111), binding energies and potential energy curves derived from DFT calculations. The trained force field was then used to study 
moir\'{e} superlattice and the corrugation of graphene.
In this Supplementary Material the technical details of the
parameter fitting procedure are shown.
In particular, we discuss the least square fitting of the
interfacial C-Cu Tersoff potential which is suitable for
simulating the gr/Cu(111) weakly bound complex.
%\end{abstract}

%\maketitle

\subsection{The Abell-Tersoff potential}

 The Abell-Tersoff (AT) potential \cite{Abell,Tersoff,Brenner,Liang} is given in the following form:
\be
V_{Tersoff}=\sum_{ij,i>j} f_{ij}(r_{ij})[V_{ij}^R (r_{ij})-b_{ij}(\Theta) V^A_{ij}(r_{ij})].
\ee
The radial part of the AT potential is composed of the
following repulsive and attractive functions,
\be
V_{ij}^R=A_{ij} exp(-\lambda_{1,ij} r_{ij}),
\ee
\be
V_{ij}^A=B_{ij} exp(-\lambda_{2,ij} r_{ij}).
\ee
The angular dependence is introduced via the attractive part $V_{ij}^A$ term by
the $b_{ij}(\Theta)$.
\be
b_{ij}(\Theta)=(1+\beta^n \chi_{ij}^n(\Theta))^{\frac{1}{2n}}
%\nonumber
\ee
%}

%\fbox{
\be
\chi_{ij}(\Theta)=\sum_{k(\neq i,j)} f_{ik}^c(r_{ik}) g_{ik}(\Theta_{ijk})  exp[\lambda_3 (r_{ij}-r_{ik})]
%\nonumber
\ee
where the cutoff function is
\[
f_{ik}(r_{ik})= \left\{ \begin{array}{cc}
%{r@{\quad: \quad}1}
~~~~~~~~~ 1 & r \le R_c-D_c  \\ \frac{1}{2}-\frac{1}{2}sin[\frac{\pi}{2}(r-R_c)/D_c] & |r-R_c| \le D_c
 \\
0 & r \ge R_c+D_c
\end{array} \right. \]
$R_c$ $\hbox{\AA}$ is the cutoff distance
and $D_c$ $\hbox{\AA}$ is the damping distance.

\begin{figure*}[hbtp]
\begin{center}
\includegraphics*[height=6cm,width=8cm,angle=0.]{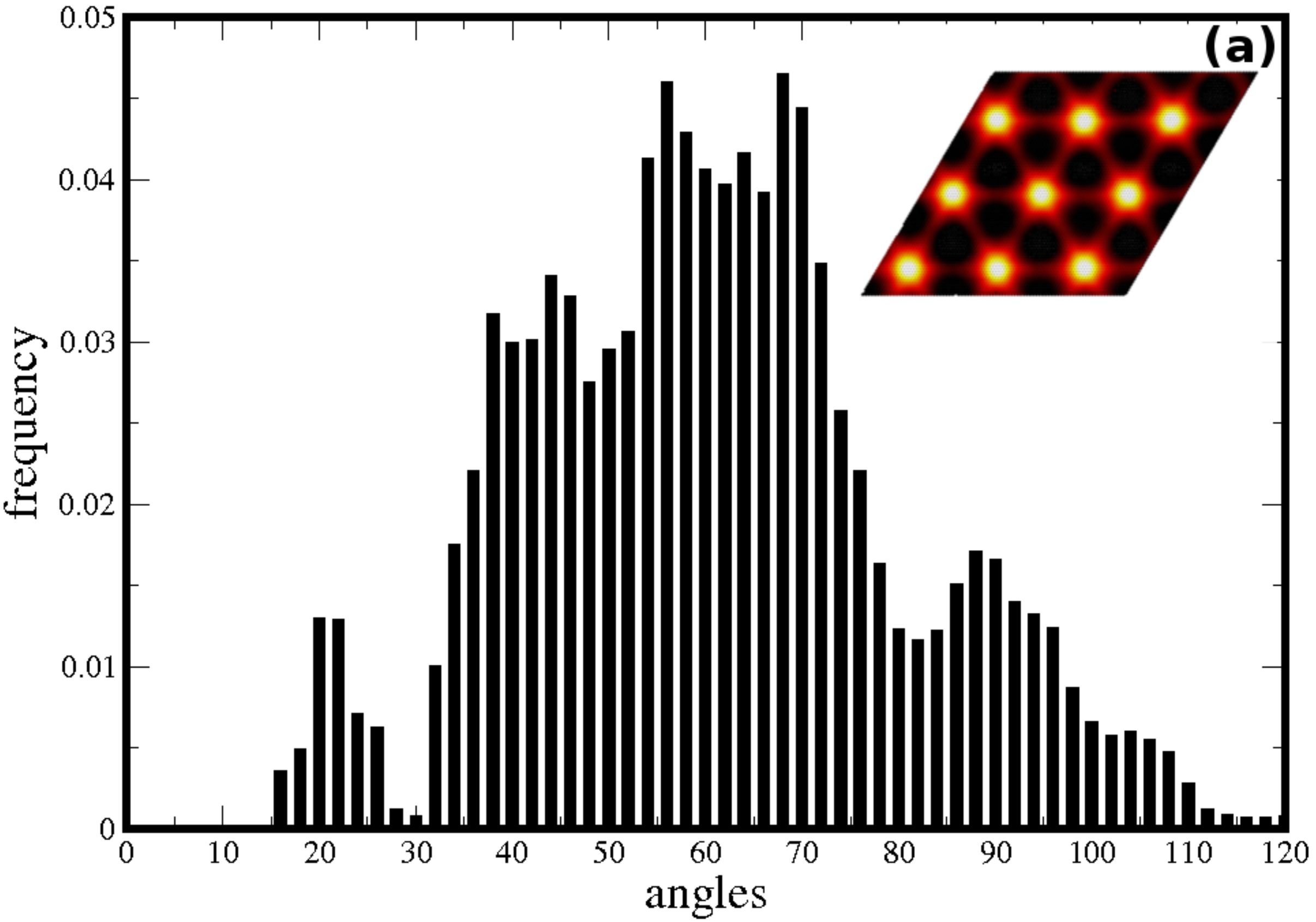}
\includegraphics*[height=6cm,width=8cm,angle=0.]{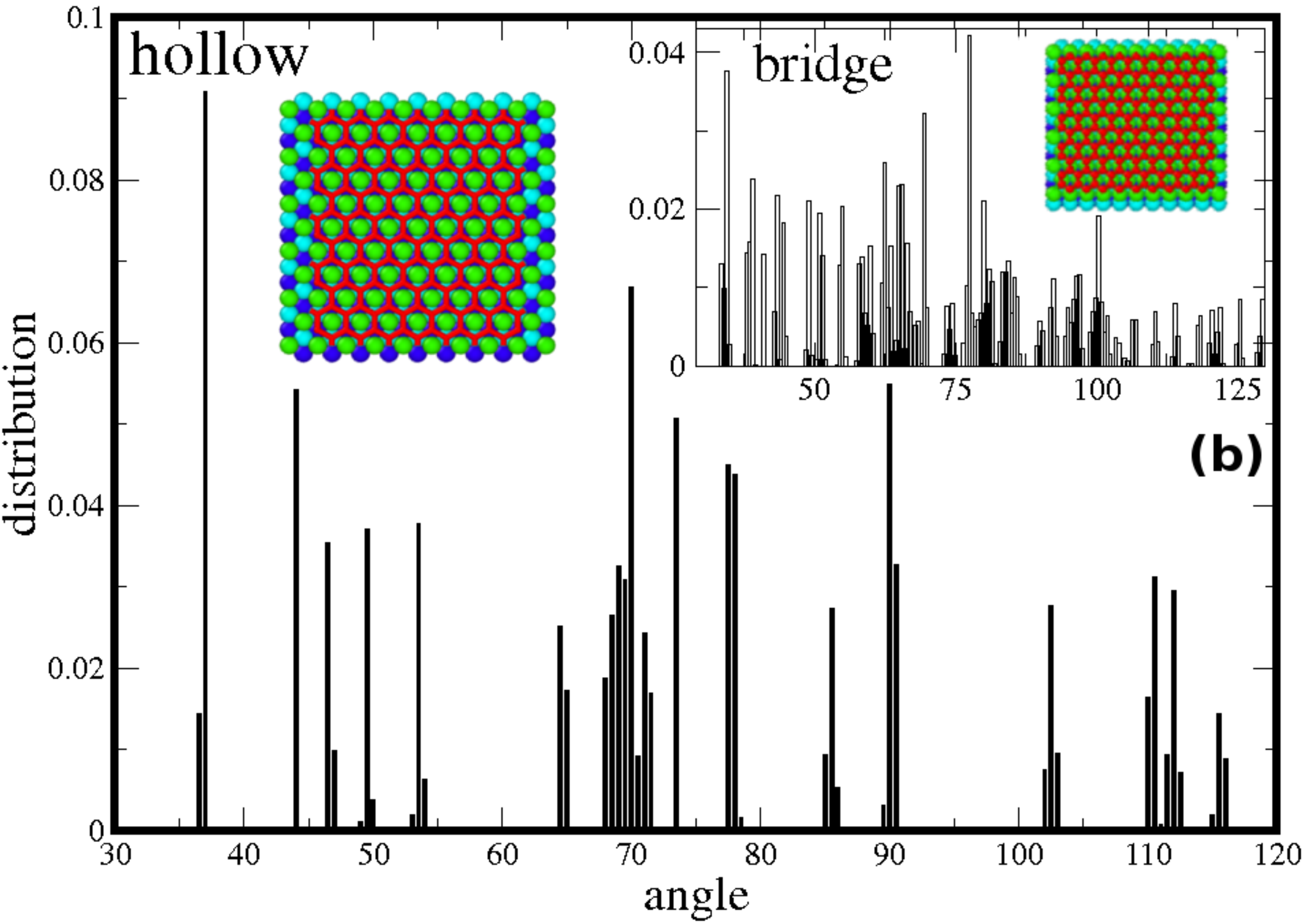}
\caption[]{
The histogram of angular distribution of the interface in 
various graphene/Cu(111) systems.
All types of interfacial angles are included (CCuC, CuCCu, CCCu, CuCuC)
and plotted against their probability.
(a) Large corrugated rhombic system with moir\'{e} pattern ($3 \times 3$ supercell)
as obtained by TL-CMD at 10 k steps.
Note the smaller peak at $\sim 25^{\circ}$ which is the lower bound of
acute out-of-plane angles (CCuC, CuCCu).
The best choice is proved to be $\Theta_0 \approx 25^{\circ}$ as found
by our least square fitting procedure.
(b) hollow configuration:
Inset: angular histogram for the flat bridge system.
}
\label{Fs1}
\end{center}
\end{figure*}

 The angular term $g(\Theta)$,
%\fbox{
\be
g(\Theta)=\gamma \biggm(1+\frac{c^2}{d^2}-\frac{c^2}{d^2+(cos\Theta-h)^2}\biggm),
%\nonumber
\ee
%}
%\\
%\vspace{0.5cm}
where $h=cos(\Theta_0)$. 
In a typical gr/Cu(111) system the interfacial bond angles (CCCu, CuCuC, CCuC,
CuCCu) varies in a wide range of $[25^{\circ};130^{\circ}]$ within cutoff
distance of the interatomic distances. This is especially true if the cutoff distance
is chosen to be relatively long-ranged. In our case $R_c+D_c \approx 4.7$
$\hbox{\AA}$, which is already long enough to account for 
van der Waals interactions. 
A typical angular distribution is shown in Fig. \ref{Fs1}(a) as obtained for
the corrugated moir\'{e} patterned superlattice.
In Fig. \ref{Fs1}(b) the angular histogram is shown for a purely hollow
modell system used for parameter fitting and for DFT calculations.
The occurrence of various angles is much less denser
than in the $3 \times 3$ superstructure.
This is partly due to the flattnes of the hollow system, but
mostly to the special angular orientation of the hollow registry.
This orientation favors specific bond angles (more discretized spectrum).
In particular, acute angles in the range of [$30^{\circ};55^{\circ}]$
also appears here.
This is important in that sense that our fitting code finds rather low $\Theta_0 \approx 25^{\circ}$
due to the acute angles in the hollow bound region.

 The main difficulty in the fitting of the Tersoff function is that
 it is almost impossible to fully account for
the precise angular distribution within the mean field approach incorporated into the $g(\Theta)$ function in the bond order term $b_{ij}(\Theta)$
in which each C-Cu heteronuclear bond angles
are treated with the same parameters.
Therefore to find a proper value for $\Theta_0$ which physically also makes sense is challenging.
We noticed that the exclusion of the in-plane bond angles 
(CCCu, CCuCu), in which the first two atoms of the atom triad in the angle
are in the gr or topmost Cu(111) layer (angles with in-plane bonds), simplifies the problem.
These angles do not influence significantly the orientation
of the interface being much softer bond angles than the
out-of-plane ("improper") angles (C-Cu-C, Cu-C-Cu).
%The latter ones can be described by 
%acute angles of $\Theta_0 \approx 30^{\circ} \pm 10^{\circ}$.
%Acute angles can occur for these angles because of the much less
%(1,3) repulsion than in e.g. CCCu and also naturally
%two nearly parallel sheets provide geomtrically acute angles.

 It turned out that these angles (out-of-plane angles) responsible for the proper orientation
of the interface: the lack of the proper choice of $\Theta_0$ results in
the weakening of the angular dependence and
leads to nanomesh-like topology instead of the required hump-and-bump like topography \cite{Sule13}.
Typical nanomesh topography occurs in nature
e.g. in h-BN/Rh(111) \cite{hbn}.
%%%%%%%%%%%%%%%%%%%%%%%%%%%%%%
% F1
%%%%%%%%%%%%%%%%%%%%%%%%%%%%%
%------------------------------------------------------
The nanomesh moir\'{e} superstructure includes
hollow-bumps (bulged-in regions) and ontop-humps (protrusions) and always provided by simple pair-potentials such as the Lennard-Jones interface potential \cite{Sule13}. 
Setting in proper angular dependence in the interface potential (e.g. the Tersoff bond order potential) the nanomesh topography
turns into an inverted topology in which the nanomesh pores
become hollow-humps and ontop-humps fall into bumps.
%Therefore, the topology is the mirror image of that of the nanomesh.
The topographic features are organized as rhombic arrays 
of protrusions.

 We also find that
other angular dependent potentials such as the Stillinger-Weber \cite{sw} or
simple pairwise+harmonic hookean angular term-like potentials do not provide proper topography
and always fall into nanomesh.
This is because these force-fields are not bond order potentials (BOPs).
Only BOPs in the form of Eq. (1) can handle adequately 
the bonding environment \cite{Tersoff} via the angular-dependent $b_{ij}(\Theta)$
bond order term.
The proper choice of the form of $b_{ij}(\Theta)$ sets in the 
proper 3-body interaction in the Abell-Tersoff potential \cite{Abell,Tersoff,Brenner,Liang}.
The bond order for instance drops for weak interactions strongly influencing
the angular orientation for such 3-body interactions.
The in-plane bond angles (e.g. CCCu) in this respect also differ from the out-of-plane
angles (e.g. C-Cu-C) since the latter ones do not include 1st order chemical bonds.
Hence the short-ranged C-C or Cu-Cu bond order terms dominate $b_{ij}(\Theta)$ for in-plane bond angles
which makes these angles much less angular dependent.
In the out-of-plane
angles the long-ranged weak C-Cu bonds are much more 
sensitive to angular fluctuations hence these angles drive
angular orientation of the gr/Cu(111) interface.
In particular, we confirmed this behavior by fitting a Tersoff function 
with a polarized bond angle dependence and no significant 
dependence of the topography on the onset of in-plane angles
has been found.
 In practice this means that the averaged contribution of the
$exp[\lambda_3 (r_{ij}-r_{ik})]$ term is negligible for
in-plane angles with respect to that of the out-of-plane ones.
This is because the $(r_{ij}-r_{ik})$ bond distance difference
fluctuates more strongly for C-Cu-C angles than for C-C-Cu atomic triads
at the interface.

 After many trial fitting procedures and simulations
we find that $h=cos \Theta_0=0.90505$ provides the best moir\'{e} pattern.
This is a somewhat surprising value if we take a look at
%\ref{} Fig. 1 in which the angular distribution is shown in a large flat and corrugated
Fig. ~\ref{Fs1} in which the angular distribution is shown in a large flat and corrugated
system. The optimal interfacial angle occurs in the wide range of $\Theta \approx 60^{\circ}
\pm 30^{\circ}$.
However, attempts were failed either with the apparent physical choice of $h \approx cos \Theta_0 \approx 0.5$ ($\Theta_0\approx 60^{\circ}$)
or with $h \approx cos \Theta_0 \approx 0.0$ ($\Theta_0\approx 90^{\circ}$)
which did not lead to perfect pattern. The selection of the less physical
acute angle of $\Theta_0\approx 25^{\circ}$ gives the best pattern.
The fitting procedure leads to this small angle without any constrain
in the parameter space.
%Although, setting $h=0.5$ and further optimization of the parameters
%gives $h=0.521280$, however, the corresponding pattern is less acceptable
%being the corrugation is too large ($\xi \approx 1.2$ $\hbox{\AA}$) (see Fig. 2).
%Also the shape and morphology of the moir\'{e} hills suffer from
%certain distorsions: in the middle a small bump appears instead of a peak.
%Interestingly, the morphology of the moir\'{e} pattern itself is not sensitive to the
%choice of $h$ and within a wide range appears ($0.9<h<-0.9$).
%However, the corrugation remains below $0.5$ $\hbox{\AA}$ only for
%$h < 0.3$, therefore we selected this range of $\Theta_0$.

\subsection{The fitting procedure}

%%%%%%%%%%%%%%%%%%%
%\subsection{Parameter fitting}
%%%%%%%%%%%%%%%%%%%

% Parameter fitting has been carried out in a similar way as described
%in ref. \cite{Sule13}.
 We used typical small representative gr/Cu(111) configurations (with flat gr)
for binding registries of hollow, top-fcc,  top-hcp
and bridge alignments (see Figs. \ref{Fs3}(c)-(f)). The
potential energy curve (PEC) of the rigid gr-Cu(111) sheet-to-sheet separation 
has been calculated by nonlocal VdW-DFT \cite{LMKLL} using the SIESTA code \cite{SIESTA}.
Then using a code developed by us \cite{potfit} the interfacial Tersoff potential
has been fitted to these DFT PECs.
   A Levenberg-Marquardt least-squares algorithm has been implemented
in the code potfit \cite{potfit} to find a combination of parameters which minimizes the deviation
between the properties in the fitting database and the properties predicted by the Tersoff potential.

 The employed properties are the followings: DFT sheet-to-sheet distance at the interface, DFT adhesion energy of the corresponding structure.
Moreover DFT PECs are used for fitting.
Using these essential 3 properties per configurations, we were able to
obtain a DFT-adaptive force field for the interface.
 The fitting
database can include various structures, although more than 2-4 modell structures
not only slows down parameter fitting but also leads to a less DFT adaptive force field (FF).
Our primary purpose was to develop a nearly perfectly macthed FF 
to suitably chosen PECs and the corresponding representative structures.
After successful parameter fitting further conditions must be
satisified by the new FF which can be tested only by test MD simulations.

 The following conditions had to be satisfied by the new FF:
(i) minimal rhomboid supercell edge size ($d \approx 6.1$ nm) for flat aligned gr
(ii) proper topology of the gr-surface: hump-and-bump morphology with a corrugation of $\xi \approx 0.4 \pm 0.1$ $\hbox{\AA}$.
Hollow-humps (moir\'{e} hills) and ontop-bumps (wells) are required 
as it has been found in other gr/substrate systems (e.g. gr/Ru(0001) \cite{Sule13}).
(iii) interface energy: adsorption or adhesion energy $E_{adh} \approx 0.11 \pm 0.05$ eV/atom
(iv) correct interfacial distances: $d_{C-Cu} \approx 3.1 \pm 0.1$ $\hbox{\AA}$.

 Among these requirements we imposed directly only
conditions (iii) and (iv) under parameter fitting.
However, the new parameter set also satisfied automatically conditions (i)-(ii).
Conditions (iii)-(iv) seem to be sufficiently strict to restrict
the parameter space in order to account for the morphology, structure
and energetics of the moir\'{e} superstructures of gr/Cu(111).
The obtained parameter set is shown in Table \ref{Ts1}.
%Further details of parameter fitting will be published in a forthcomming paper
%\cite{Sule2}.

 The fitting procedure has been carried out in a few steps.
\\
{\em Step (1)}: First an initial guess of radial parameters have been obtained
using the receipt given by Albe and Erhart \cite{SiC,GaN}.
%In particular, $D_0$ and $r_0$ can be estimated from the adhesion energy of gr/Cu(111) ($E_{adh} \approx 0.11$ eV/C) and equilibrium distance of gr-Cu(111) ($r_0 \approx 3.1$ $\hbox{\AA}$).
Then using formulas in ref. \cite{Sule13} one can estimate the initial guess for $A$,
$B$, $\lambda_1$ and $\lambda_2$.
\\
%{\em Step (2)}: Then using the code pontifix \cite{pontifix} 
%the entire parameter space has been optimized using a training set
%with few general unit cells of weakly bound alloys representing
%the bonding of C-Ru alloys.
\\
{\em Step (2)}: Finally, the obtained parameter set
has been refined by an additional code written in our laboratory
\cite{potfit}.
In this case we consider {\em ab initio} DFT potential energy curves and/or equilibrium
DFT geometries of small gr/Cu(111) modell systems. 
Using this way of parameter fitting we were able to
get an adequate force field which describe gr/Cu(111) interfacial bonding properly.

 The traditional way of fitting procedure (see e.g. refs. \cite{SiC,GaN})
does not work in this special case when a weak interface potential is to be
parameterized.
In a standard situation one should fit the Tersoff function to the
experimental lattice constants, cohesive energies and bulk moduli of various polymorphs
of CuC.
However, in this case the interface potential would bind graphene
too strongly to Cu(111) (chemical adhesion).
The bonding situation and the chemical environment is completely
different in gr/Cu(111) and in CuC.
Even if a weak chemical bonding takes place in gr/Cu(0001), it is far much weaker
than in CuC.
Using a CuC based fitted potential
the adhesion energy of gr/Cu(111) would be $E_{adh} \gg 1$ eV/C, which is far higher than
the measured and the DFT calculated $E_{adh} \approx 0.1$ eV/C \cite{grcu_adhesion}.
%Moreover, in the CuC dimer molecule, e.g. the dissociation energy is in the range of
%6.6 eV \cite{CuCdim} which is far above the adhesion energy of
%gr/Cu(111) of 0.1 eV/C \cite{}, hence the
%dimer properties nor can be used for fitting.
%Moreover, the equilibrium distance of the CuC dimer is also too short ($1.66$ $\hbox{\AA}$
%when compared with the expected $\sim 3.1$ $\hbox{\AA}$ distance of the
%gr/Cu(111) complex.
Hence one can not use the available experimental data set of CuC for parameterization.

 Instead we directly fitted the free parameters in the Tersoff expression
on a training set of small configurations of gr/Cu(111) in a similar way
as it has been done for gr/Ru(0001) in ref. \cite{Sule13}.
The only difference is that we do not employ here the code PONTIFIX \cite{pontifix}, instead we estimate the initial guess of the parameters as
described above in step (1).
 In step (2) we used our code for fitting the parameters
on this realistic data base.
%This final training set include small gr/Cu(111)
%interaction using the SIESTA code \cite{siesta}.

%%%%%%%%%%%%%%%%%%%%%%%%%%%%%%
% F2
%%%%%%%%%%%%%%%%%%%%%%%%%%%%%
\begin{figure*}[hbtp]
\begin{center}
\includegraphics*[height=6cm,width=8cm,angle=0.]{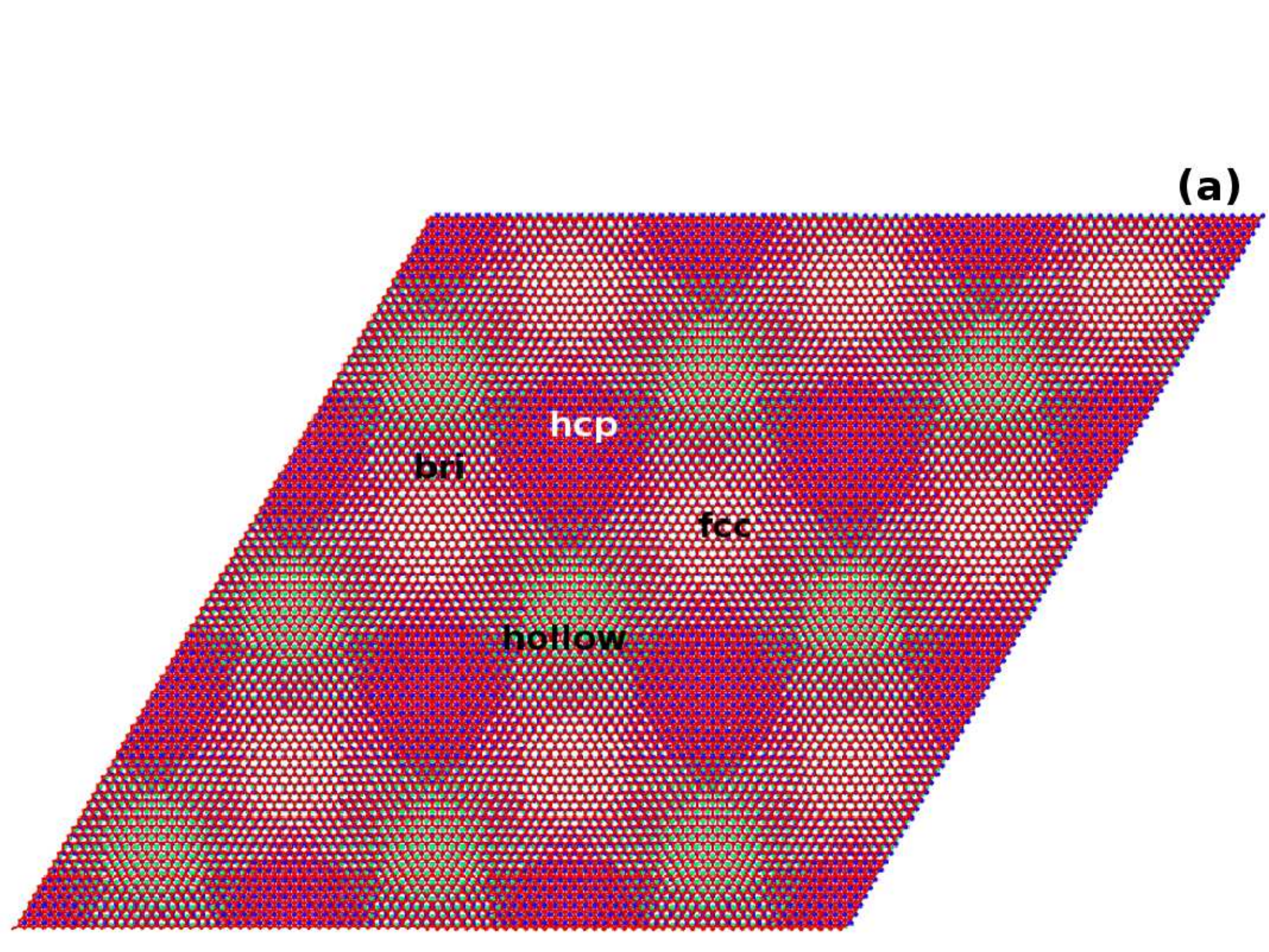}
\includegraphics*[height=6cm,width=8cm,angle=0.]{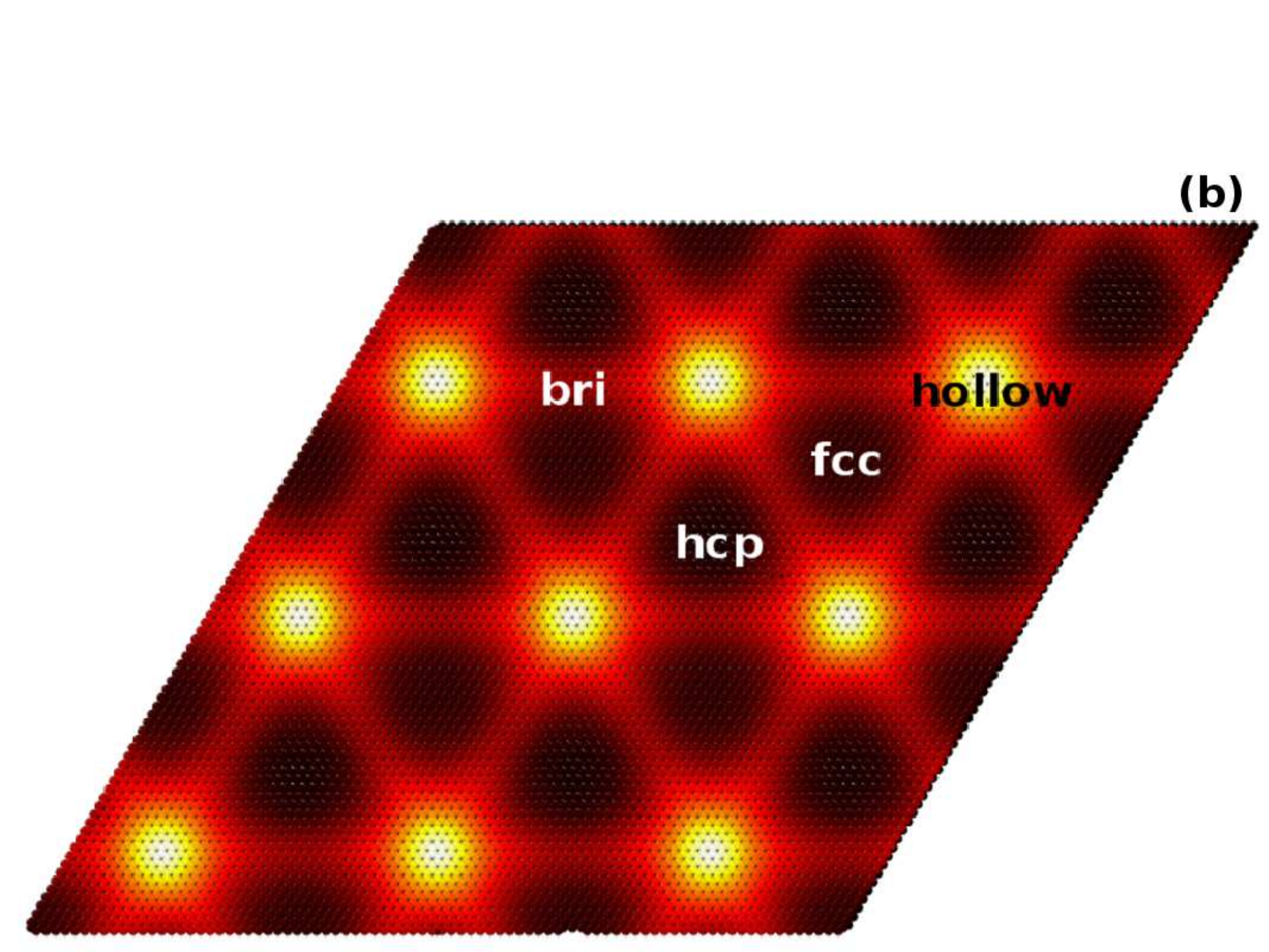}
\caption[]{
(a) The flat rhomboid $3 \times 3$ superlattice before geometry optimization (rigid lattice) which is used for the PEC calculation in Fig. \ref{Fs3}(a)
including 27898 total atoms (with 3 Cu topmost layers).
The various binding sites are denoted and assigned using the same
notations as shown e.g. in ref. \cite{Batzill} (see Fig. 3, p. 86).
(b) The time averaged moir\'{e} superlattice at 0.1 K with
the assigned binding registries.
}
\label{Fs2}
\end{center}
\end{figure*}
%------------------------------------------------------

 Additional requirements are the following (besides mentioned already above): the average corrugation is being below $0.5$ $\hbox{\AA}$ even at 300 K,
moreover, the 0 K structure should be stable at 300 K with
minor corrugation increase,
minimal C-Cu distance $d_{min} > 2.9$ $\hbox{\AA}$,
maximal C-Cu distance $d_{max} < 4.5$ $\hbox{\AA}$,
no decorations occurs on the surface besides the regular hexagonally shaped humps
(no further protrusions, vacancy islands, or holes).

%%%%%%%%%%%%%%%%%%%%%%%%%%%%%%%%%%%%%%%
% T1 Table 1
%%%%%%%%%%%%%%%%%%%%%%%%%%%%%%%%%%%%%%%
%\LARGE{
\begin{table}[t]
\caption[]
{
The fitted Abell-Tersoff parameters for the graphene/Cu(111) interface.
}
\label{Ts1}
\begin{ruledtabular}
\begin{tabular}{lc}
%\rowcolor[gray]{.8} C-Cu (Tersoff, present work)  & weak & strong    \\
C-Cu (Tersoff)  &      \\
\hline%
%%%%%%%%%%%%%%%%%%%%%%%%%%%%%%%%%%%%%%%%%%%%%%%%%%%%%%%%
%Cu C C 1.0000000000000000 0.0883167511417047 1.5527865760271629 40.9755961701796707 0.9528753276972811 0.9050528410835819 1.0000000000000000 1.0000000000000000 2.0455965425671860 320.7794950040236017 4.1978604878268992 0.4794477186682328 3.1308174157727371 977.7958178882481661
% #   m, gamma, lambda3, c, d, costheta0, n,
% #   beta, lambda2, B, R, D, lambda1, A
%%%%%%%%%%%%%%%%%%%%%%%%%%%%%%%%%%%%%%%%%%%%%%%%%%%%%%%%
A (eV)      & 977.795817888248 \\
B (eV)      & 320.779495004024 \\
$\lambda_1$ & 3.13081741577273  \\
$\lambda_2$ & 2.04559654256718      \\
$\gamma$    & 0.08831675114170 \\
c           & 40.9755961701790    \\
d           & 0.95287532769728  \\
h           & 0.90505284108358    \\
$R_c$ ($\hbox{\AA}$)         & 4.197860487827   \\
$D_c$ ($\hbox{\AA}$)         & 0.479447718668      \\
$\beta$ ($\hbox{\AA}^{-1}$)  & 1.0    \\
%$\mu$                        & 1         \\
$\lambda_3$                  & 1.552786576027 \\
n,m                          & 1   \\
\hline
%---------------------------------------------------------------
\end{tabular}
\end{ruledtabular}
%}
\footnotetext[1]{
%\footnotemark[1]*{
%pw: present work,
The parameters have been fitted to small flat gr/Cu(111) systems.
Notations are the same as used in ref. \cite{Sule13} (supplementary material)
and on the web page of lammps \cite{lammps}.
$\lambda_3$ is denoted as $\mu$ and $h=cos \Theta_0$ in the supplementary material of ref. \cite{Sule13}.
}
\end{table}

%%%%%%%%%%%%%%%%%%%%%%%%%%%%%%
% F3
%%%%%%%%%%%%%%%%%%%%%%%%%%%%%
\begin{figure*}[hbtp]
\begin{center}
\includegraphics*[height=7cm,width=9cm,angle=0.]{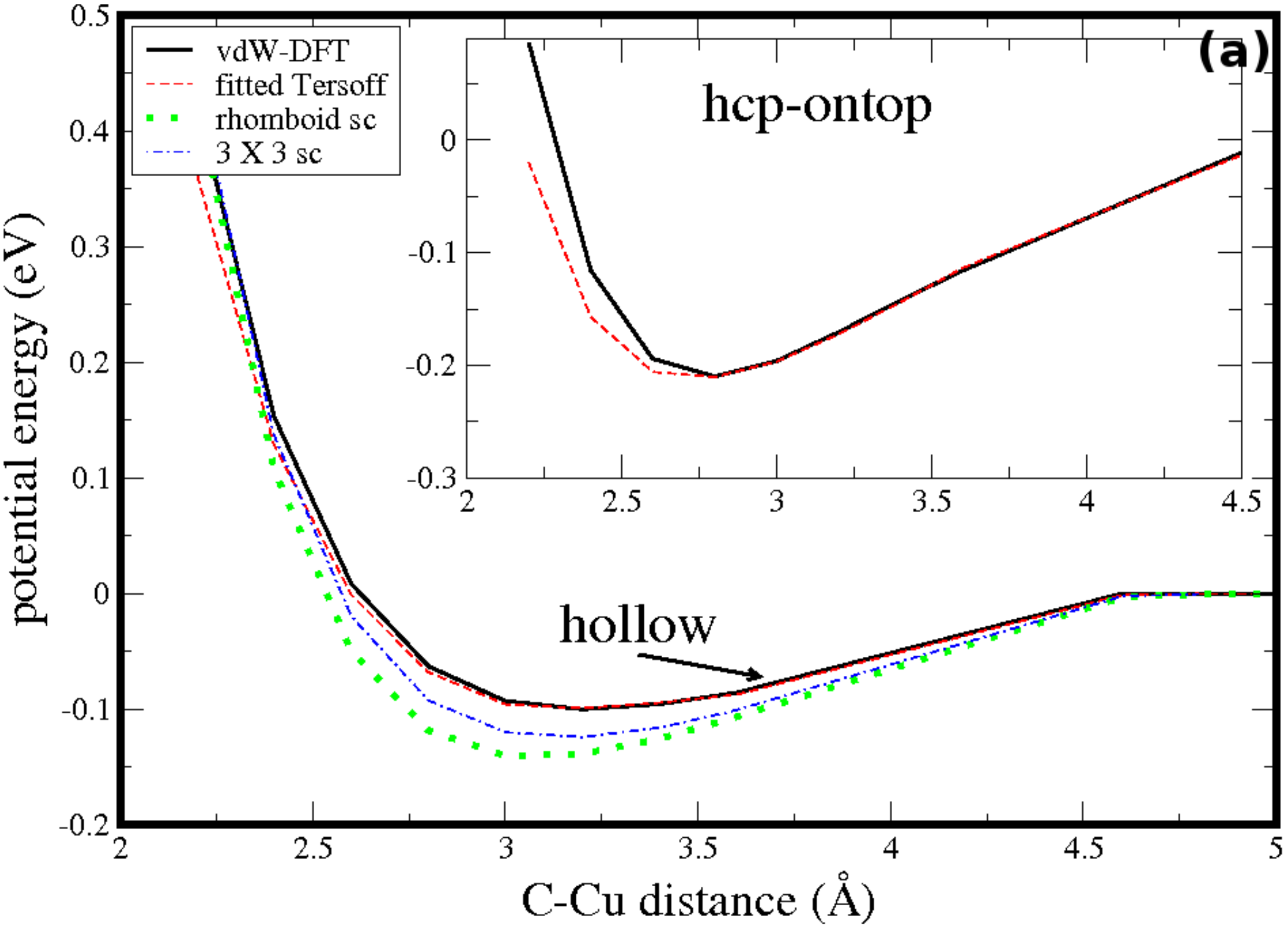}
\includegraphics*[height=7cm,width=8cm,angle=0.]{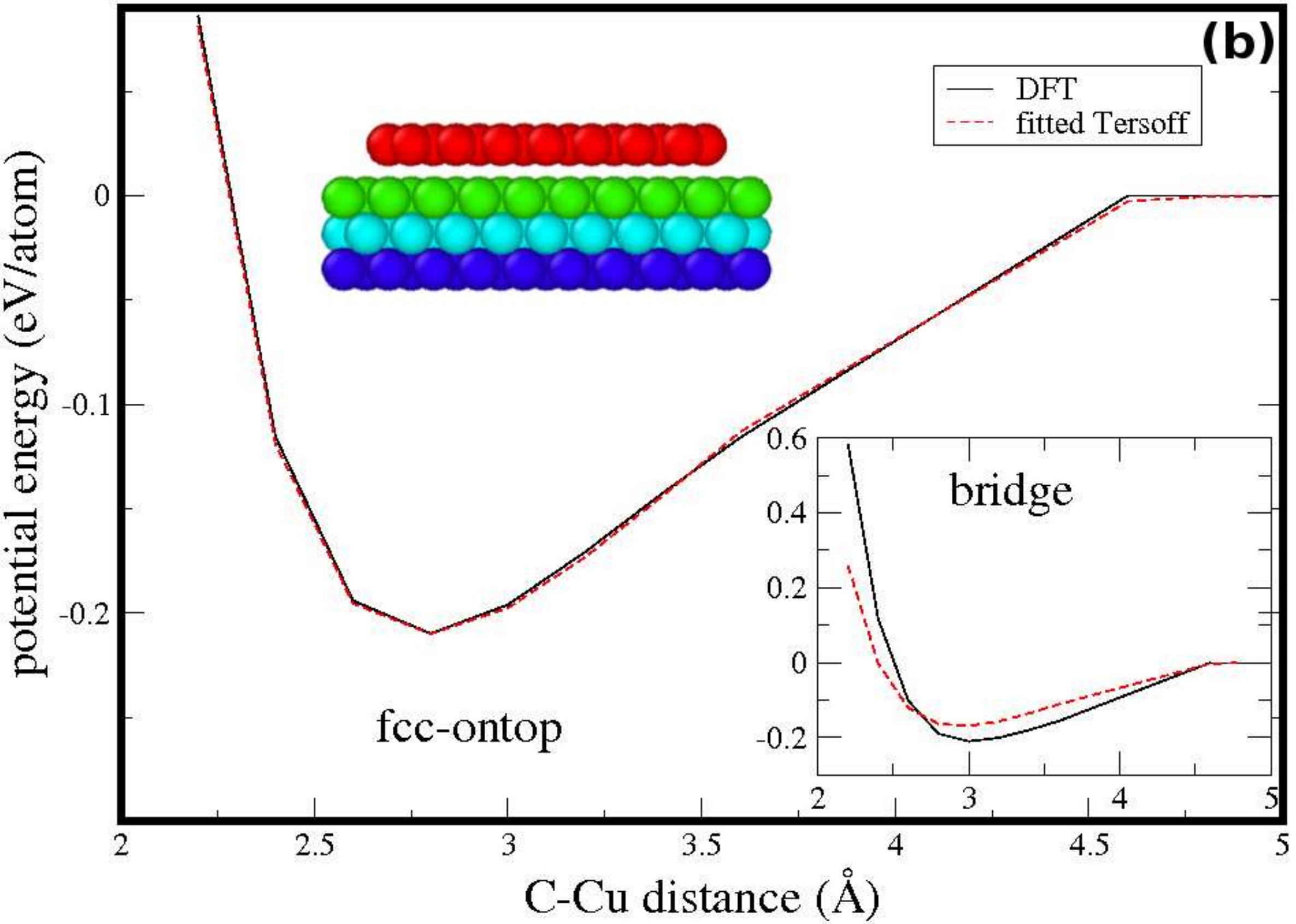}
\\
\includegraphics*[height=6cm,width=8.5cm,angle=0.]{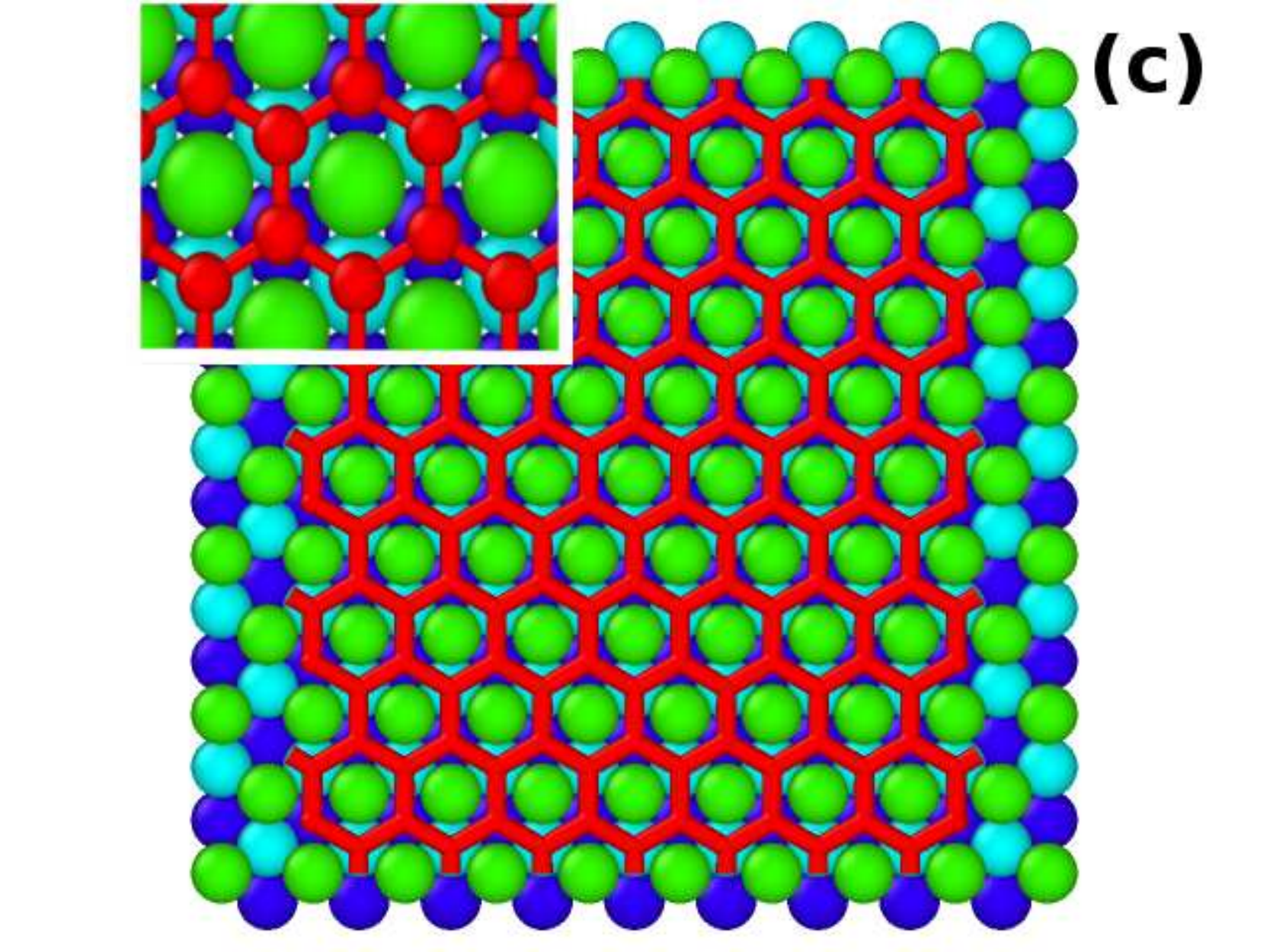}
\includegraphics*[height=6cm,width=8cm,angle=0.]{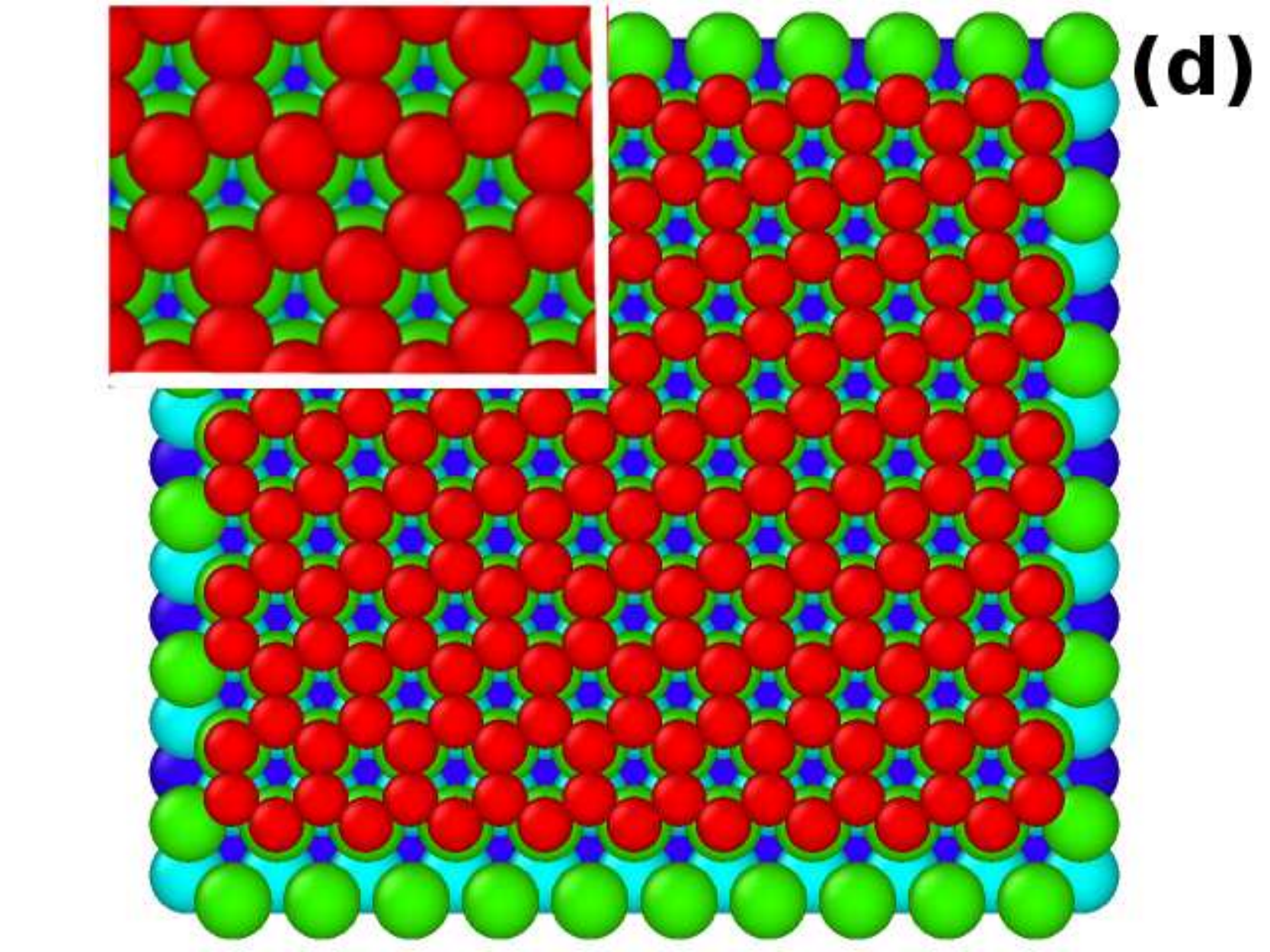}
\includegraphics*[height=6cm,width=8cm,angle=0.]{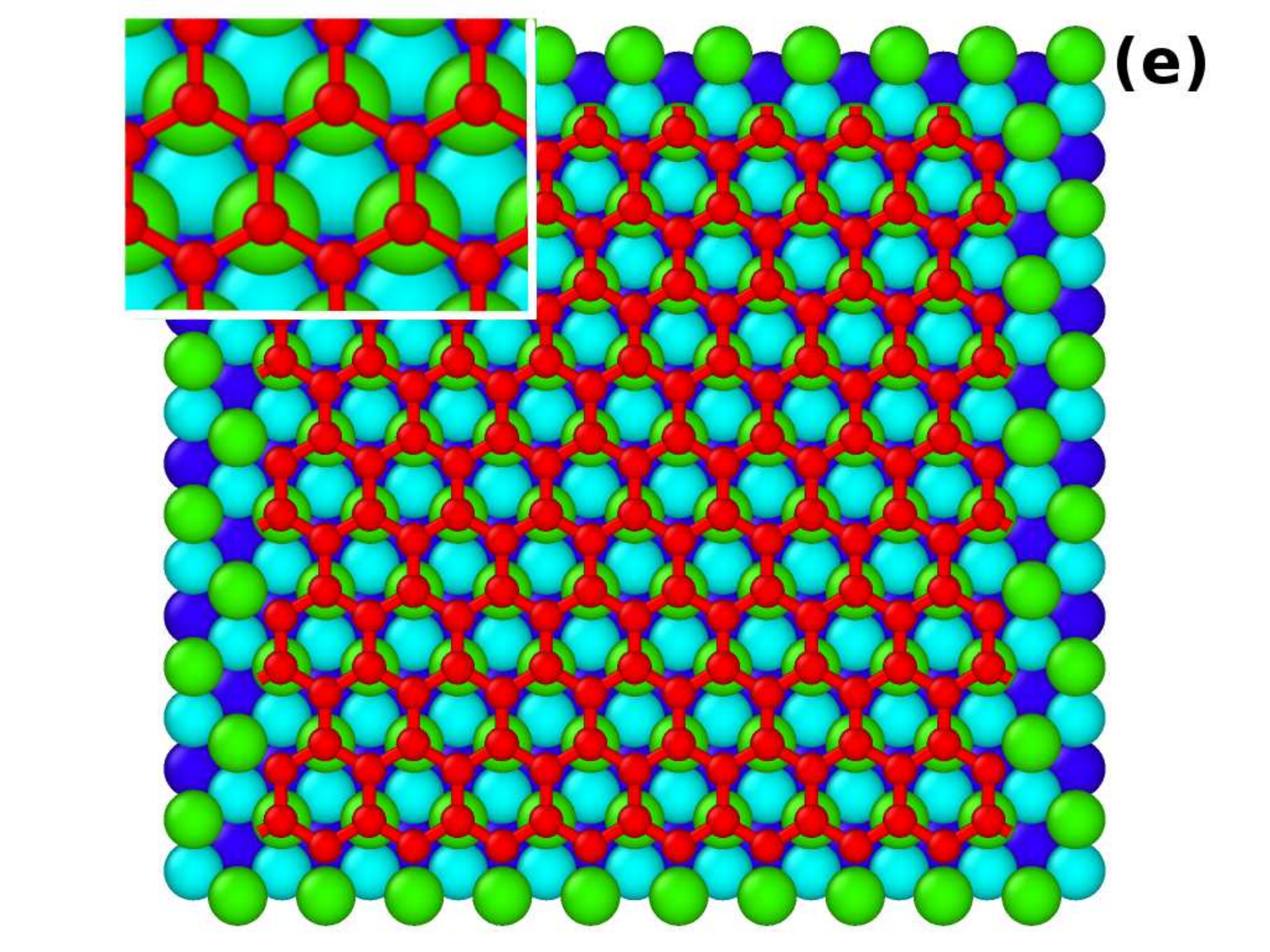}
\includegraphics*[height=6cm,width=8cm,angle=0.]{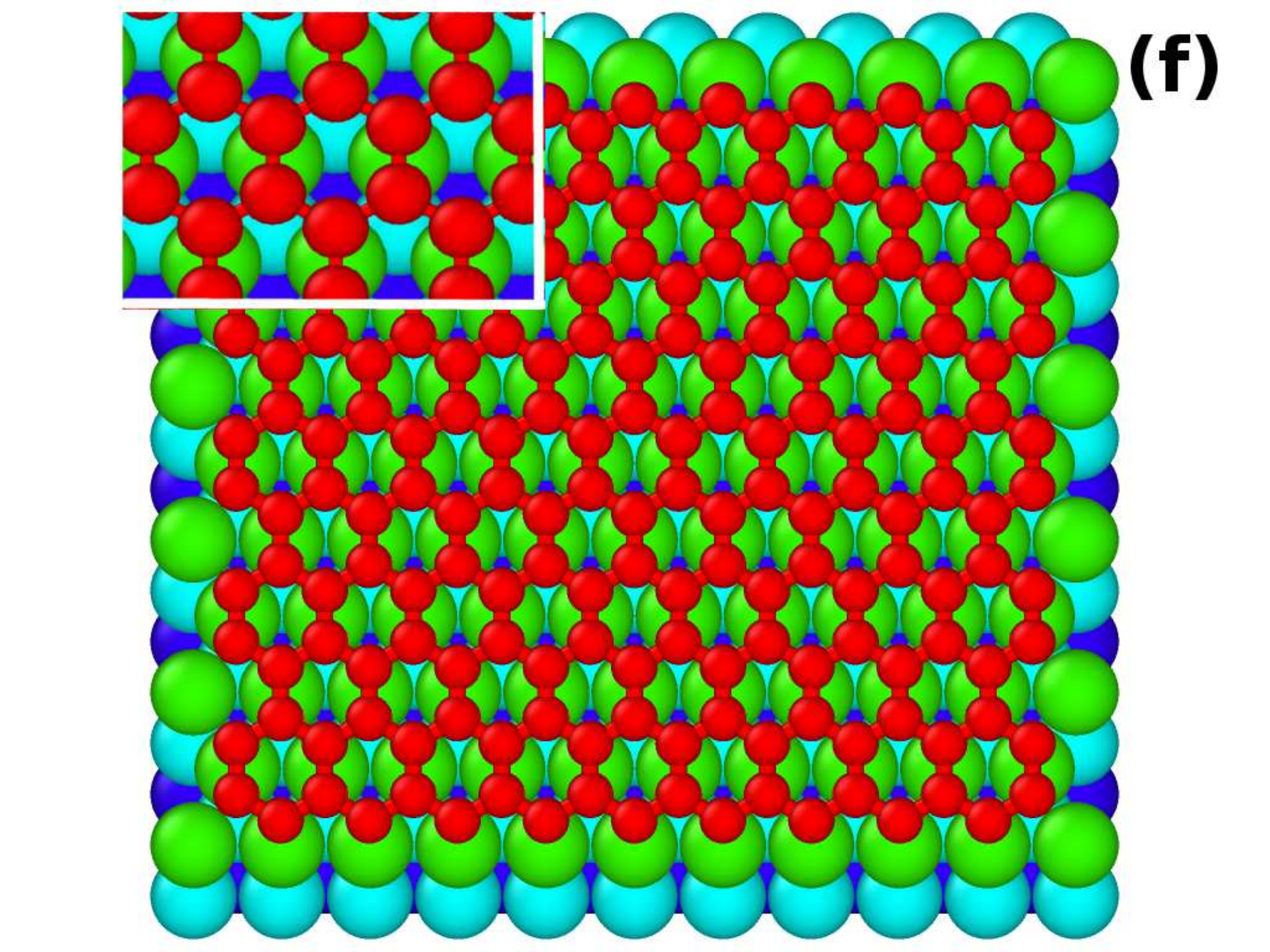}
\caption[]{
(a) The potential energy curves (PECs) as a function of the 
sheet-to-topmost Cu(111) layer distance (C-Cu) obtained for flat graphene/Cu(111)
systems by the nonlocal vdW-DFT method and by the fitted Abell-Tersoff
potential developed in this paper.
The potential energy curve for two different configurations with different binding registries
are shown: nearly purely hollow system (Carbon atoms in hollow positions)
and an ontop system with hcp registries (Inset Fig. \ref{Fs3}a).
The gr sheet has been expanded to match the appropriate lattice positions
of the Cu(111) support.
The PEC is also shown for flat larger systems with 3183 (rhomb sc) and 27898 ($3 \times 3$ sc) atoms (the minimal rhomboid supercell with the two topmost layers of Cu and the large $3 \times 3$ superlattice) as calculated by the new
Tersoff potential (no DFT PEC is available for these large systems).
The following small modell systems have been used in the training set:
The hollow (hcpfcc) (c) and the hcp-ontop (tophcp) (d) system used for fitting and
for DFT calculations (the training set).
The bridge (e) and fcc (topfcc) (f) registries are not included in the training set.
The hollow, atop-fcc, atop-hcp and bridge notations are the same
as it was given in ref. \cite{Batzill} (see Fig. 1, p. 86). 
}
\label{Fs3}
\end{center}
\end{figure*}
%------------------------------------------------------

\subsection{The fitting database}

For fitting we used small representative gr/Cu(111) systems which are suitable
for DFT calculations. These are typically smaller than 1000 atoms.
We also separated different binding registry systems. 
We found it important because one has to force the free parameters 
to account correctly for weaker hollow adherence and for
stronger ontop adhesion. 
In our fitting code \cite{potfit} we were able to select in the fitting database
and we could use different weights for different systems depending
on the required results.  
In Figs \ref{Fs3}(a) and \ref{Fs3}(b) the obtained potential energy curves can be seen as
obtained by a nonlocal vdW-DFT method \cite{LMKLL}
and by the fitted Tersoff potential for hollow, hcp-ontop (a) and for fcc, bridge (b) configurations.
It can clearly be seen that in the hollow position adherence is much weaker,
roughly the half of that of the hcp position.
In general, the interfacial Tersoff potential follows the DFT curve
and alters slightly from it mostly for the hcp and bridge systems at short C-Cu atomic separations 
(repulsive part 
of the curve).
This could be the reason that the adhesion energy is slightly overestimated.
The bridge system, which was not included in the training set,
shows up some deviation from the DFT curve.
Including, however, the bridge registry system in the fitting data base
leads to the further deepening of the adhesion energy and to the
deterioration of the match to the other DFT PECs.
After many trials we found the hcp-hollow systems are suitable for fitting.

 The potential energy curve (PEC) as a function of the C-Cu (sheet-to-sheet) distance
has also been calculated by the fitted Tersoff potential
for flat rhomboid supercells of different sizes.
The corresponding PECs are also shown in Figs. \ref{Fs3}(a) and \ref{Fs3}(b).
Increasing the size of the systems the PEC is getting closer to
the PEC of the hollow system.
Although for these large systems no DFT data is available, however,
the calculated Tersoff potential provides PEC rather similar in shape
to that of the hollow system
and the well depth and position are also in the same range.
The larger $3 \times 3$ supercell is a mixed system with various
binding sites (hollow, hcp, fcc and bridge, see Fig. \ref{Fs2}(b))).
The corresponding PEC in Fig. \ref{Fs3}(a) reports us that
the PEC is, however, ruled by the hollow registry,
although the more strongly bound ontop sites (hcp, fcc and bridge)
have also significant contribution to the average PEC.

\end{document}